\documentclass[useAMS,usenatbib]{mnras}
\usepackage{graphicx}
\usepackage{amsmath}
\usepackage{amssymb}
\usepackage{color}
\usepackage{hyperref}

\title[Multiple kinematic structures in a NIHAO Milky Way]{Introducing {\tt galactic structure finder}: the multiple stellar kinematic structures of a simulated Milky Way mass galaxy}
 \author[Obreja et al]{Aura Obreja$^{1,2}$\thanks{E-mail: obreja@usm.lmu.de}, Andrea V. Macci\`{o}$^{2,3}$, Benjamin Moster$^{1,4}$, Aaron A. Dutton$^{2}$, 
 \and Tobias Buck$^{3}$, Gregory S. Stinson and Liang Wang$^{5}$\\   
 $^{1}$University Observatory Munich, Scheinerstr. 1, D-81679 Munich, Germany\\
 $^{2}$New York University Abu Dhabi, PO Box 129188, Saadiyat Island, Abu Dhabi, United Arab Emirates\\
 $^{3}$Max Planck Institute f\"{u}r Astronomie, K\"{o}nigstuhl 17, 69117 Heidelberg, Germany\\
 $^{4}$Max-Planck-Institut f\"{u}r Astrophysik, Karl-Schwarzschild Stra\ss e 1, 85748 Garching, Germany\\
 $^{5}$University of Western Australia, Crawley, WA 6009, Australia}

\date{Accepted 2018 April 18. Received 2018 February 16; in original form 2017 December 18} 
 
\begin{document}

\maketitle

\label{firstpage}

\begin{abstract}
We present the first results of applying Gaussian Mixture Models in the stellar kinematic space of normalized angular momentum and binding energy 
on NIHAO high resolution galaxies to separate the stars into multiple components.  
We exemplify this method using a simulated Milky Way analogue, whose stellar component hosts: thin and thick discs, classical and pseudo bulges, and a stellar halo.  
The properties of these stellar structures are in good agreement with observational expectations in terms of sizes, shapes and rotational support. 
Interestingly, the two kinematic discs show surface mass density profiles more centrally concentrated than exponentials, while the bulges and the stellar halo are purely exponential. 
We trace back in time the Lagrangian mass of each component separately to study their formation history. 
Between $z\sim3$ and the end of halo virialization, $z\sim1.3$, all components lose a fraction of their angular momentum. 
The classical bulge loses the most ($\sim95\%$) and the thin disc the least ($\sim60\%$). 
Both bulges formed their stars in-situ at high redshift,  while the thin disc formed $\sim98\%$ in-situ, but with a constant
$SFR\sim1.5M_{\rm\odot}yr^{\rm-1}$ over the last $\rm\sim11$~Gyr.  
Accreted stars ($6\%$ of total stellar mass) are mainly incorporated to the thick disc or the stellar halo,  
which formed ex-situ $8\%$ and $45\%$ of their respective masses. 
Our analysis pipeline is freely available at \url{https://github.com/aobr/gsf}.
\end{abstract}


\section{Introduction}

The classification of observed galaxies into distinct types has a long history \citep{Sandage:1961}. 
It is much more than an abstract taxonomic exercise with many implications for studies of poorly understood physical phenomena. 
Two notorious examples are: the relation between supermassive black holes (SMBHs) and their host galaxies, and the determination of the dark matter (DM) halo profiles. 
Currently, the general consensus is that SMBH masses correlate tightly with the stellar velocity dispersion of their host bulges \citep{Ferrarese:2000,Gebhardt:2000,Haring:2004}, 
implying that the two most probably co--evolve \citep[but see also][]{Jahnke:2011}.  
Also the kinematics of galaxy discs have a long history of being used to infer the underlaying dark matter distributions \citep{Rubin:1980}. 
Hence, an accurate theory on the formation of galactic stellar structures like discs and bulges would invariably lead to tighter constraints on these puzzling phenomena, 
apart from filling in the gaps on galaxy formation models in general. 

While in observations these structures are defined by various luminosity profiles fitted to galaxy images \citep[e.g.][]{vanderWel:2012}, 
as today there is yet no clear link between them and the intrinsic, kinematically defined stellar structures. 
A way to bridge this gap is provided by high resolution galaxy simulations which have full information on the stellar phase space (position and velocities),
thus allowing for a proper definition of the intrinsic kinematic structures. Simulations can also be post--processed with radiative transfer codes
to create mock images that can be subsequently analyzed as it is done for galaxy observations \citep[e.g.][]{Obreja:2014,Guidi:2016,Buck:2017,Bottrell:2017}. 
In this manner, it is in principle possible to search for a quantitative relation between the photometric morphology of a galaxy and its stellar kinematic structures. 
Therefore, it is necessary to have the means of robustly defining galactic stellar kinematic structures in simulations, which is precisely the aim of this work.

In \citet[][hereafter Paper I]{Obreja:2016} we showed how dynamic stellar discs can be separated from the spheroids in galaxy simulations by using  
a clustering algorithm in a multidimensional kinematic space. However, galaxies are known to sometimes host a much larger variety of stellar structures.   
The stars of the Milky Way, in particular, are though to form several components: a thin and a thick disc \citep{Gilmore:1983}, a boxy/peanut bulge \citep{Okuda:1977}, 
a nuclear star cluster \citep{Becklin:1968}, a bar \citep{Hammersley:2000} and a stellar halo \citep{Searle:1978}. 
The evidences for a classical bulge in the Milky Way are currently still debated \citep[e.g.][]{Bland-Hawthorn:2016}.
In extragalactic studies, many of the nearby spirals seen close to edge-on are better described by a two component disc rather than by one 
\citep{Dalcanton:2002,Yoachim:2006,Comeron:2011,Comeron:2014,Elmegreen:2017}. The inner regions of observed galaxies also appear to sometimes host multiple bulges and/or bars
\citep[e.g.][]{Athanassoula:2005,Gadotti:2009,Aguerri:2009,Nowak:2010,Kormendy:2010,Mendez-Abreu:2014}.
All these observational data encouraged us to improve the method described in Paper I to be able to disentangle more then two kinematic components in simulations. 

In the current paper we present the result of applying this method to a simulated Milky Way, hereafter MW, mass galaxy from the NIHAO sample \citep{Wang:2015}.  
We have done the same analysis for a total of 25 NIHAO galaxies, ranging from dwarfs to galaxies a few times more massive than the Milky Way, and found various combinations of stellar structures.
For this paper, however, we chose one galaxy which resembles the Milky Way in a few important aspects, to exemplify how our pipeline works. 
The results on the complete 25 galaxy sample are the subject of an accompanying work \citep{Obreja:2018}.

The particular galaxy we use as test case turns out to have five stellar kinematic components: a thin and a thick disc, a classical and a pseudo bulge, and a stellar halo, 
with properties within the expected observational ranges for shapes, velocities, rotational support and specific angular momenta. 
We also study the evolution of these properties for the five components separately to learn more about their formation patterns. 

In recent years cosmological simulations have started to achieve enough resolution to make it possible to study galactic stellar structures
\citep[e.g.][]{Scannapieco:2010,Brook:2012, Aumer:2013, Stinson:2013a, Christensen:2014, Hopkins:2014, Marinacci:2014, Schaye:2015, Wang:2015, Grand:2017}.
In this light, we make our analysis code freely available with the hope it 
will provide the means for a self consistent study of the formation and evolution of such structures 
across different simulation codes with different numerical schemes and feedback implementations.

This work is structured as follows. Section~\ref{methods} presents the method to search for stellar structures. 
The simulated galaxy we are using as a test is described in Section~\ref{sim_section}.
The results of our method applied to this galaxy are analyzed and discussed in Section~\ref{g8.26e11}.
Section~\ref{z0prop_mw} gives the properties of the stellar kinematic structures at redshift $z=0$ 
in comparison to the Milky Way, while Section~\ref{z0prop} presents the results of analysing the galaxy from an extragalactic point of view. 
The evolution of the kinematic structures is given in Section~\ref{evolution}. 
Finally we summarize our results and highlight some concluding remarks in Section~\ref{conclusions}.

\begin{figure} 
\includegraphics[width=0.47\textwidth]{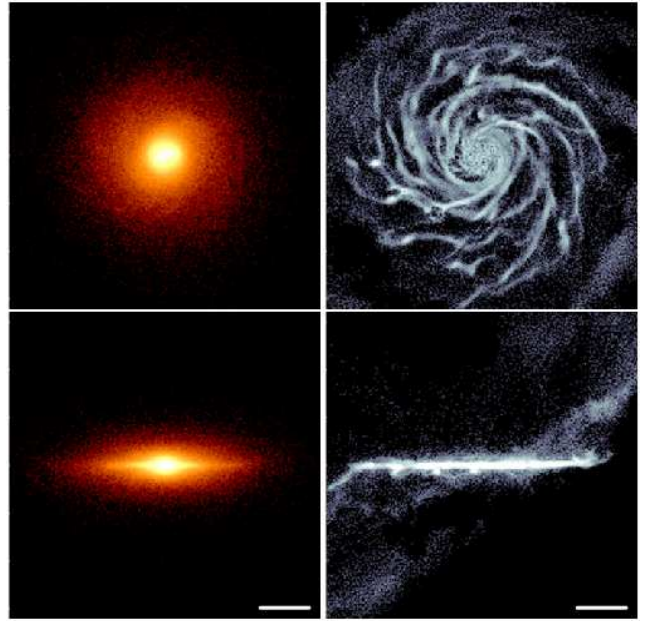} 
\caption{The surface mass density maps for the stars (left) and the cold gas (right) in face-on (top) and edge-on (bottom) projections for the galaxy g8.26e11.
The white horizontal lines represent the physical scale of 10 kpc.}
\label{fig:sunrise}
\end{figure}

\section{Gaussian Mixture Models applied to galaxy dynamics}
\label{methods}

The most widely employed method to kinematically split simulated galaxies was introduced by \citet{Abadi:2003}, 
and is based on the distribution of stellar circularities  
\citep[e.g.][]{Brooks:2008,Scannapieco:2010,Scannapieco:2011,Brook:2012,Martig:2012,Kannan:2015,Grand:2017}. 
The circularity parameter $\epsilon$ is the ratio between the azimuthal angular momentum of a particle $J_z$, 
and the angular momentum of a circular orbit having the same binding energy $J_c(E)$, 
where the $z$-direction is along the symmetry axis of the galaxy.

\citet{Domenech:2012} proposed another method which uses not only the circularity parameter, but also the 
binding energy of the particles $E$, and the angular momentum component $J_p$ defined as $\overrightarrow{J}_p=\overrightarrow{J}-\overrightarrow{J}_z$, 
and normalized also to the angular momentum of the circular orbit. Here $\overrightarrow{J}$ is the total angular momentum of a stellar particle.  

These authors use the \textit{k-means} cluster finding algorithm \citep{Scholkopf:1998} with a given number of groups to disentangle the 
stellar galaxy components in this 3D space, ($J_{\rm z}/J_{\rm c}(E)$, $J_{\rm p}/_{\rm c}(E)$, $E$). 
The \textit{k-means} algorithm minimizes the sum of squared distances for all cluster pairs, also known as intra-cluster distance 
or cluster ``inertia''. This method needs to assume a certain metric, and even though it will always converge given 
enough iterations, it might be to a local minimum. The main limitations of \textit{k-means} are the assumptions of cluster 
convexity and isotropy. Therefore,  \textit{k-means} is best suited for regular shaped manifolds and spherical 
clusters that are approximately equally populated. 

In Paper I we generalized the method of \citet{Domenech:2012}
by using \textit{Gaussian Mixture Models} (hereafter GMM) instead of \textit{k-means}, in a similar 3D dynamical space.
GMM is a probabilistic method that results in a so called soft assignment of particles to clusters, 
each particles having a normalized probability to belong to a certain group.
Same as the \textit{k-means}, it is an iterative method which employs
an expectation--maximisation algorithm to find the parameters of the Gaussians. 
This method relaxes the assumption of cluster symmetry, allowing for a fully free covariance matrix.
Also, since it uses the Mahalanobis distance to the cluster centers (means of the Gaussians) as the minimization criteria, 
it does not bias the results towards equally weighted clusters. 
On the particular problem of separating the dynamical components of a galaxy's stars, it naturally results in 
analogues of observed substructures like thin and/or thick discs, classical- and/or pseudo-bulges, and/or stellar haloes,
with mass weights that are not constrained to be roughly equal. 

Throughout this study, the 3D dynamical space refers to ($j_z/j_c$, $j_p/j_c$, $e/|e|_{\rm max}$),
lower case letters denoting \emph{specific} angular momentum and binding energy.
The specific binding energies are scaled to the absolute value of the energy of the most bound stellar particle in the halo $|e|_{\rm max}$. 
Therefore $-1<e/|e|_{\rm max}<0$, and as such the galaxy/dark matter halo mass dependence is factored out, as well as the dimensionality of the energy.
A crucial step between the previous and the present study is that the halo potential is now recomputed assuming isolation. 
In this manner pathological distributions of binding energy can be circumvented. 

The complete analysis package, which we call {\tt galactic structures finder} or {\tt gsf}, can be downloaded via \url{https://github.com/aobr/gsf}. 
It is a Python-Fortran90 package based on {\tt pynbody} \citep[\texttt{http://pynbody.github.io},][]{Pontzen:2013} 
to load, orient and transform to physical units a simulation snapshot, 
and the {\tt scikit-learn} Python package for Machine Learning \citep{Pedregosa:2011} to run the clustering algorithm. 
An OpenMP Fortran90 module was added to perform the direct N-body gravity force using all the particles in the given halo. 
The {\tt gsf} analysis package assumes that the simulation snapshot has been pre-processed with a halo finder.
For this study we have used {\tt Amiga Halo Finder} \citep[{\tt AHF},][]{Knollmann:2009}.
The analysis package has been designed to work out of the box with simulations that can be loaded with {\tt pynbody}.

The work flow of {\tt gsf} is as follows:
\begin{itemize}
 \item The simulated halo is loaded with {\tt pynbody} and converted to physical units.
 \item The halo is oriented with the $z$-axis parallel to the galaxy's total stellar angular momentum. 
 \item The gravitational potential at the position of each stellar particle due to all (dark matter and baryons) particles in the halo is computed by direct summation.
 \item The gravitational potential in the equatorial plane at fixed radial positions is computed by direct summation over all particles in the halo.
 \item At the same radial position on the equatorial plane, the code computes the specific angular momentum of particles on circular orbits, 
 and constructs the $e-j_c$ mapping.
 \item The specific angular momenta of the stellar particles are decomposed as $\vec{j}=\vec{j_z}+\vec{j_p}$, 
 and their corresponding $j_c$ are computed by interpolating on the $e-j_c$ mapping the accurate values of their binding energies calculated previously. 
 \item The input feature matrix $(j_z/j_c, j_p/j_c, e/|e|_{\rm max})$ with as many entries as stellar particles is passed to the clustering algorithm
 together with the number of groups $nk$ to look for. 
 \item The clustering algorithm returns a matrix of probabilities $P_{ik}$, where $i$ is the stellar particle index and $k$ runs from $0$ to $nk-1$. 
 For each $i$ the probabilities are normalized: $\Sigma_{k=0,nk-1}P_{ik}=1$.
 \item The code creates two types of figures. The first one contains the stellar mass distributions in the input parameters $j_z/j_c$, $j_p/j_c$ and $e/|e|_{\rm max}$
 for each of the $nk$ structures found by the clustering algorithm. The other type of figure is made for each of the $nk$ structure separately, 
 and contains the face-on and edge-on stellar surface mass densities, and the edge-on line-of-sight velocity maps.  
\end{itemize}
All the relevant information of a run is saved into various files. This includes: the stellar indices $i$ and the matrix of probabilities $P_{ik}$, 
the re-computed gravitational potential of all stellar particles, the $e-j_c$ mapping, and the rotation matrix needed to transform the raw simulation
to the equatorial plane of the galaxy.

\begin{figure*}
\includegraphics[width=0.98\textwidth]{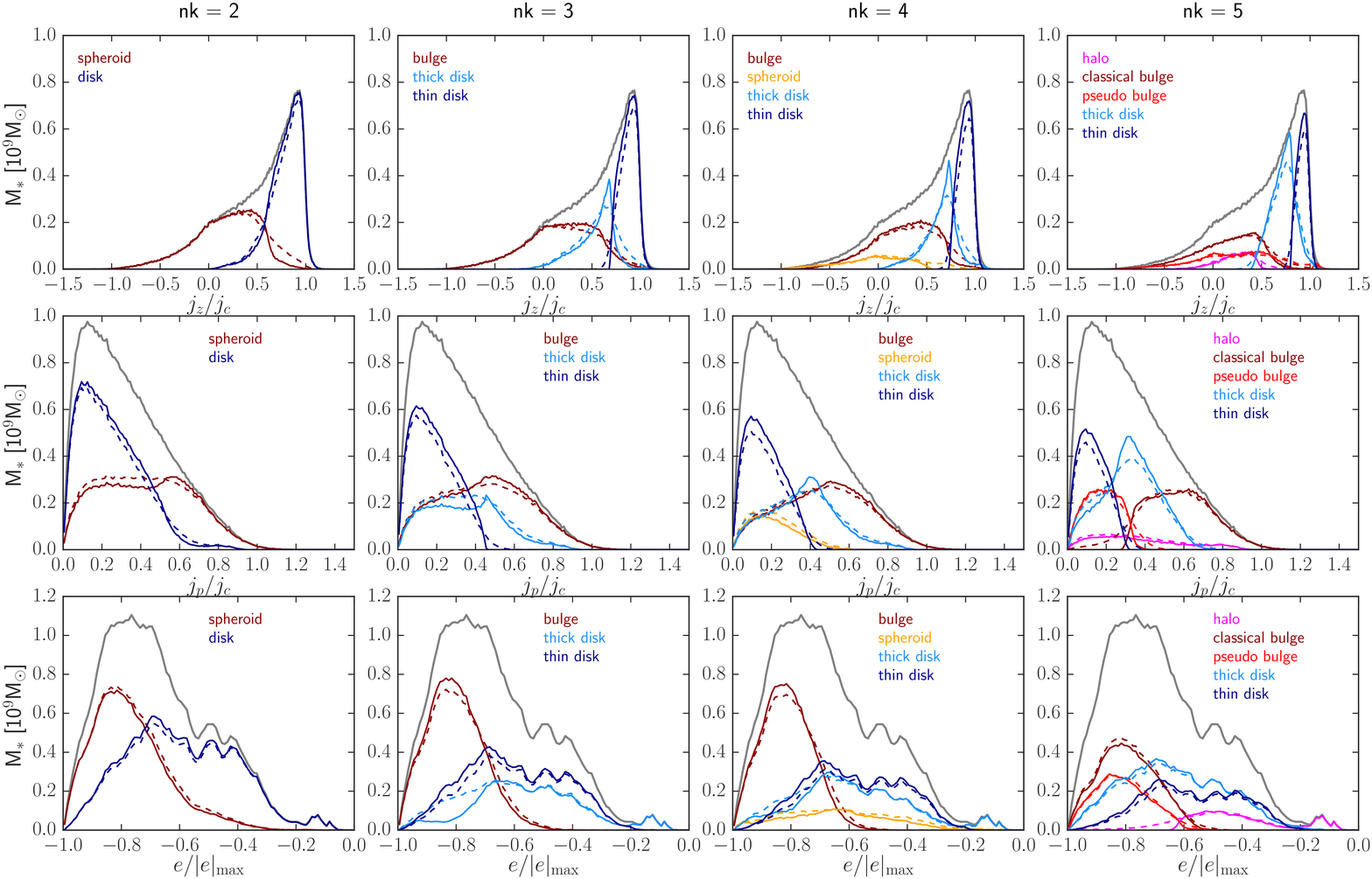}
\caption{The results of {\tt gsf} applied to the galaxy g8.26e11 shown as stellar mass in each component as a 
function of the dynamical features given as input: $j_{\rm z}/j_{\rm c}$ (top row), 
$j_{\rm p}/j_{\rm c}$ (central row) and $e/|e|_{\rm max}$ (bottom row), when varying the number of components from $nk=2$ (far left)
to $nk=5$ (far right). The solid and dashed colored lines stand for the hard and soft clustering tagging (see text for more details), 
while the solid grey give the total stellar mass. The colored labels in each panel are the components' nicknames.
For all panels the bin width is fixed to 0.01.}
\label{figure3}
\end{figure*}

\begin{figure*}
\includegraphics[width=0.98\textwidth]{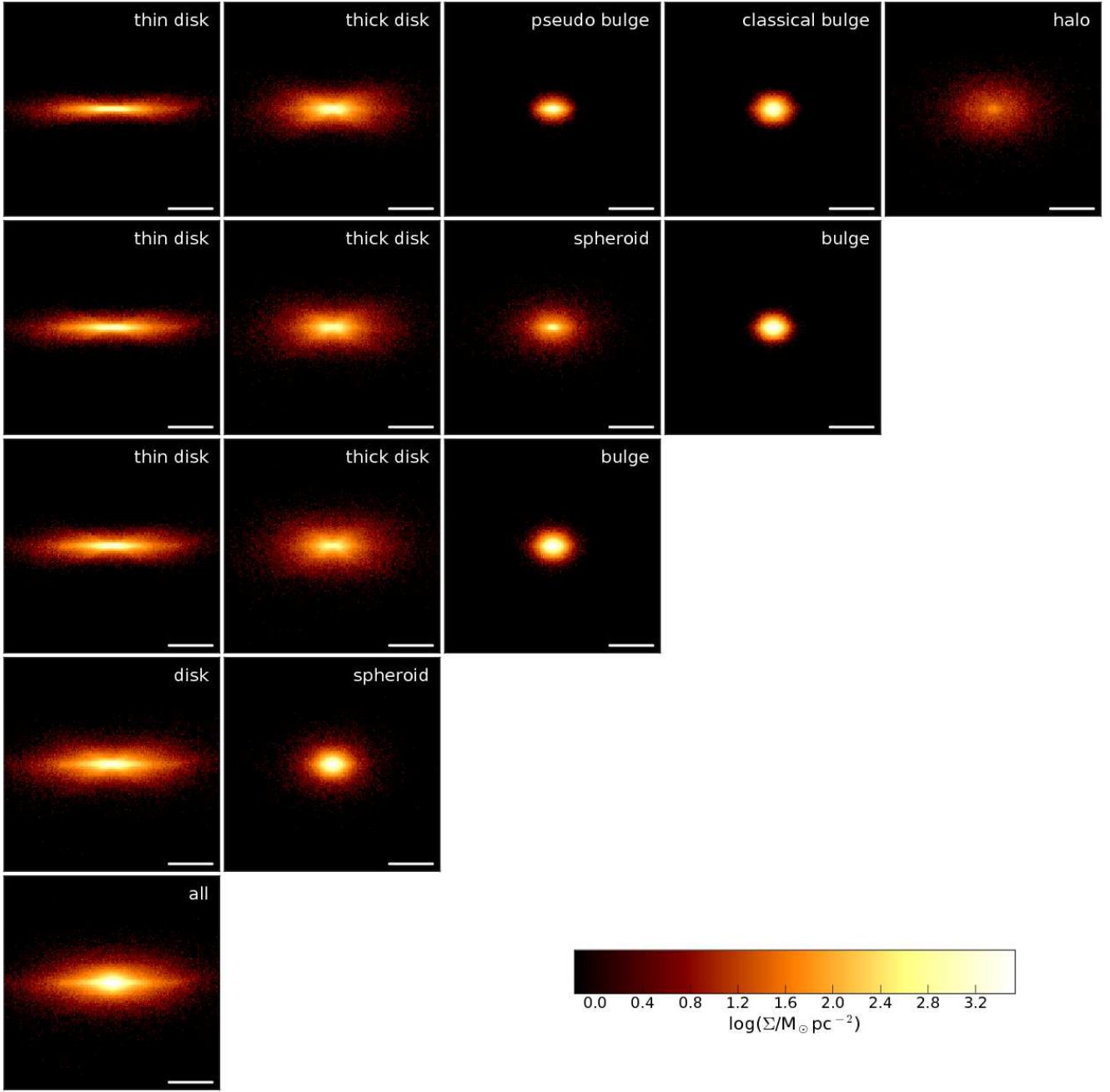}
\caption{The edge-on surface mass density of the stellar components of g8.26e11 for $nk=2$ (second row from the bottom) 
to $nk=5$ (top row). The complete edge-on surface mass density is shown in the bottom left corner. The white bars give the 10 kpc
physical scale, and the white labels provide the correspondence with the same components shown in Figure~\ref{figure3}.
The range of mass surface densities is the same for all panels.}
\label{figure4}
\end{figure*}

\begin{figure*}
\includegraphics[width=0.98\textwidth]{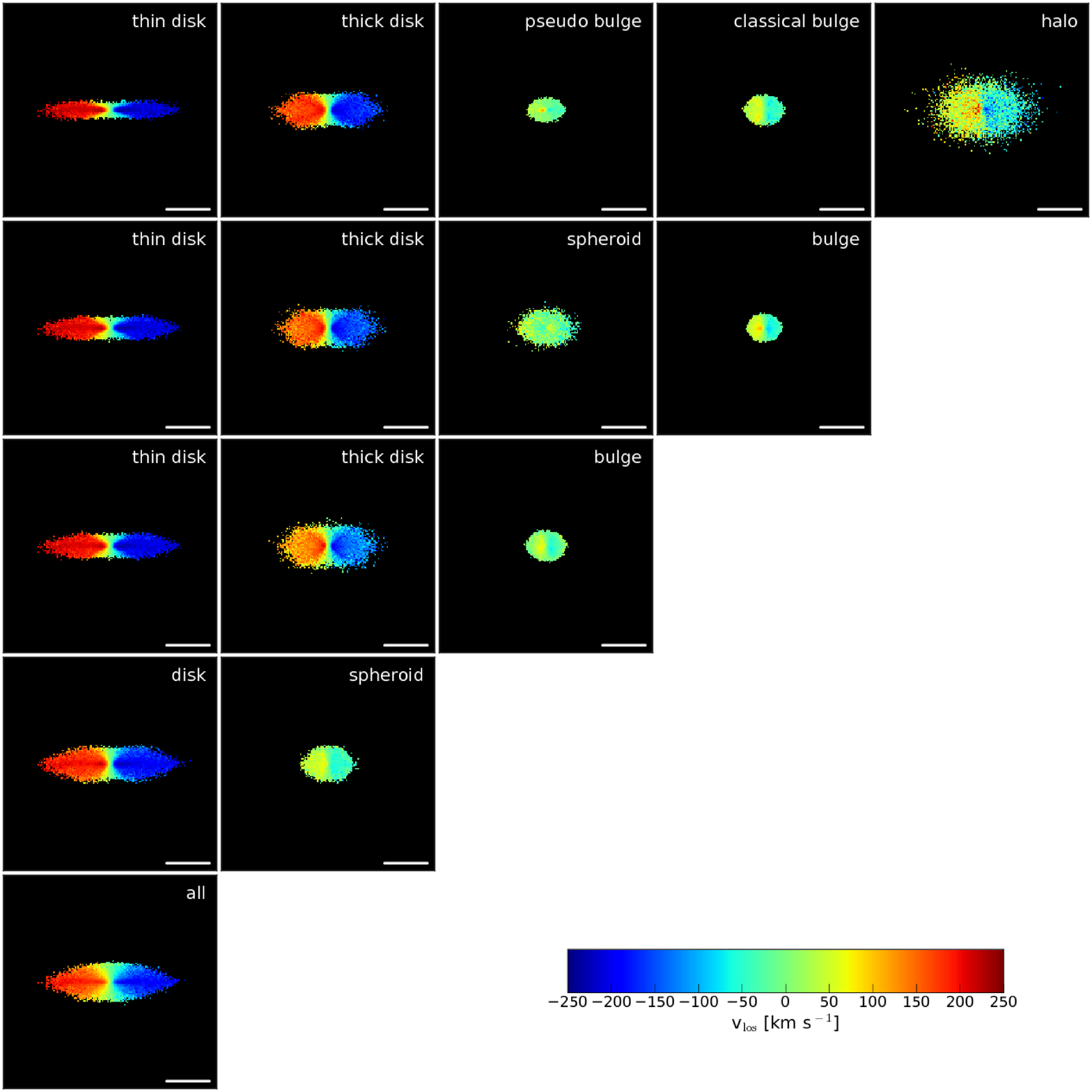}
\caption{The edge-on line of sight velocities of the stellar components of g8.26e11. 
The white bars give the 10 kpc physical scale, and the white labels provide the correspondence with the same components shown in 
Figures~\ref{figure3} and \ref{figure4}. All panels have the same velocity range.}
\label{figure4.1}
\end{figure*}

\section{The simulations}
\label{sim_section}

The NIHAO suite \citep{Wang:2015} is a series of baryonic cosmological zoom-in simulations run with the improved version \citep{Wadsley:2017} 
of the N-body SPH code {\tt GASOLINE} \citep{Wadsley:2004}, assuming a standard Planck cosmology (Planck Collaboration 2014). 
The version of the code used for the run includes fixes to deal with the artificial cold blobs \citep{Ritchie:2001} by 
employing the artificial viscosity implementation of \citet{Price:2008}. 
The SPH kernel is that of \citet{Dehnen:2012}, assuming 50 neighbors. 
In order to better resolve the shocks induced by feedback the code uses the time step limiter of \citet{Saitoh:2009}.
Metals are diffused as discussed in \citet{Wadsley:2008}. 
The sources of gas heating are photoionization and photoheating from a redshift dependent UV background \citep{Haardt:2012},
while the cooling channels are the metal lines and the Compton scattering \citep{Shen:2010}.
Gas particles with temperatures lower than 15000 K and densities higher than 10.3 cm$^{\rm -3}$ can form stars following a Kennicutt-Schmidt relation.
Stellar feedback takes into account two processes: the SNe II blast-waves \citep{Stinson:2006} and the pre-heating of the gas in the region where such a event will take place
by the massive star which is the SN II progenitor \citep{Stinson:2013a}. The implementation of the latter process is also known as ``early stellar feedback''. 
The code assumes a Chabrier Initial Mass Function \citep[IMF;][]{Chabrier:2003}. 
The heavy element enrichment of the gas is based on the SNe Ia yields of \citet{Thielemann:1986} and SNe II yields of \citet{Woosley:1995}.

The NIHAO simulations cover three orders of magnitude in dark matter halo mass, from dwarfs to galaxies at the peak of the baryon conversion efficiency. 
In this mass regime the SN feedback is supposed to be the dominant factor limiting star formation, while AGN feedback (not included in this version of {\tt GASOLINE}) should have only a marginal effect. 
All these galaxies indeed follow the redshift dependent abundance matching constraints \citep{Moster:2013,Behroozi:2013}, 
and thus can also be used to study galaxy evolution. 
For the particular purpose of studying the evolution of galactic stellar structures, we have chosen a subsample of 25 NIHAO galaxies, which mainly comprises the massive end of the complete sample. 
The reason we excluded the simulated dwarfs is that it is generally not expected of these types of galaxies to have a large variety of stellar dynamical subcomponents. 
Also, given that our method is intended to work on virialized systems, we have excluded the massive galaxies which at $z=0$ have disturbed stellar mass distributions.

The analysis presented in this work has, thus, been done on a sample of 25 simulated galaxies, 
from which we chose one galaxy (g8.26e11) to show how {\tt gsf} is capable of disentangling stellar kinematic structures with clear observational counterparts.
This particular galaxy has the total mass, stellar mass and morphology very close to the Milky Way.

The galaxy g8.26e11 has a mass resolution of 3.2$\rm\times$10$^{\rm 5}$M$_{\rm\odot}$ and 1.7$\rm\times$10$^{\rm 6}$M$_{\rm\odot}$ for the 
gas and dark matter particles, respectively. Its gravitational softenings are 400~pc and 931~pc, respectively. At $z=0$, this galaxy has a 
dark matter halo mass of 9.0$\rm\times$10$^{\rm 11}$M$_{\rm\odot}$, a stellar mass of 4.7$\rm\times$10$^{\rm 10}$M$_{\rm\odot}$, and a viral radius of 213~kpc. 
The mass of cold gas (T$\rm<$15000~K) is 4.2$\rm\times$10$^{\rm 10}$M$_{\rm\odot}$, while the virial fraction of cold gas is 0.57. 
This galaxy, same as the complete NIHAO sample, respects the Tully-Fisher \citep{Tully:1977} relations for both stars and baryons \citep{Dutton:2017}.

Figure~\ref{fig:sunrise} shows the face-on (top) and edge-on (bottom) projections for the stellar (left) and cold gas (right) surface mass densities of g8.26e11.
This galaxy resembles a large design spiral from the nearby Universe, as it can be appreciated from the face-on gas projection.

\section{The stellar kinematic structures of a Milky Way mass galaxy}
\label{g8.26e11}

We start our study by showing how increasing the number of components, $nk$, required by {\tt gsf}, the search algorithm naturally leads 
to dynamical stellar structures that can be associated with the various components thought to be part of observed galaxies, and particularly of the MW. 

Figure~\ref{figure3} shows the results of running {\tt gsf} for the galaxy g8.26e11 when $nk$ is increased from $2$ to $5$ (left to right), 
by plotting the mass in each component as a function of the input dynamical features, $j_z/j_c$, $j_p/j_c$, $e/|e|_{\rm max}$ (top to bottom). 
The different colors stand for the various components, each being given a nickname, which is also shown in the figure. 
The solid/dashed lines represent the hard/soft clustering assignations. 
The soft tagging means that each particle $i$ has a certain combination of probabilities $\{P_{k}^{(i)}\}$ to belong to the GMM groups $\{k\}$ with $k$ running from $0$ to $nk-1$, 
such that $\Sigma_k P_k^{(i)}=1$. 
The hard tagging associates each particle $i$ to the one group $k$ for which $P_k^{(i)}$ is maximum.
Therefore, to construct the solid curves each stellar particle contributes all its mass to the one group to which most likely belongs,
while for the dashed curves each particle proportionally distributes its mass to the $\{k\}$ groups according to its probabilities $\{P_k^{(i)}\}$.
The fact that the two ways of tagging are so similar can be invoked as a good reason for using the hard assignations, 
which is the case for the rest of this study.

The results in Figure~\ref{figure3} are transformed to `observables', namely edge-on mass surface densities and velocity maps in Figures~\ref{figure4} and ~\ref{figure4.1}, respectively,
where $nk$ decreases from top down. The nicknames of the various components have been chosen based on the maps in Figures~\ref{figure4} and ~\ref{figure4.1}.
This manner of choosing the name of the components is only feasible for small sample of galaxies. We are currently exploring various possibilities to perform this step 
automatically. 

The circularity histogram for this galaxy has a strong peak close to $j_{\rm z}/j_{\rm c}=1$ and no other important feature (grey curves in the top panels of Figure~\ref{figure3}).

For the run with $nk=2$, {\tt gsf} distinguishes the mass under the sharp circularity peak (dark blue) from the broad, 
more symmetric distribution centered close to $j_z/j_c=0$ (dark red), as can be appreciated from the top left panel. 
Looking at the same two components in the other two dynamical features (center and bottom left panels), 
the first component (dark blue) is obviously more biased towards orbits in the equatorial plane, with a peak in $j_p/j_c\sim0.1$ than the second one (dark red),
with an almost flat distribution between $j_p/j_c\sim0.1$ and $\sim0.8$, and less gravitationally bound (bottom left panel). 
The corresponding edge-on mass surface densities and velocity maps in the forth rows of Figures~\ref{figure4} and ~\ref{figure4.1} for the two $nk=2$ components given by the dark blue and dark red 
distributions in the left column of Figure~\ref{figure3} show the expected characteristics of \textit{discs} and \textit{spheroids}: 
axial symmetry and a spider velocity diagram for the former versus spherical symmetry and only a small amount of coherent rotation for the latter.  

Increasing $nk$ to $3$, the material of the disc component from $nk=2$ is redistributed into two new components (center left column of Figure~\ref{figure3}), 
one containing only material from the $nk=2$ disc, while the other also encloses some of the $nk=2$ spheroid. 
The new component shown in light blue gathers most of the least rotationally supported material of the $nk=2$ disc,  $j_z/j_c$ from $0$ to $\sim0.6$ 
(far left and centre left top panels), and the least gravitationally bound mass of the $nk=2$ spheroid (far left bottom panel). 
The new component in light blue of the $nk=3$ run has a large $j_p/j_c$ range $[0,0.9]$, while the corresponding component in dark blue 
is now more confined to the equatorial plane ($j_p/j_c<0.45$) than the disk of $nk=2$.
As it can be expected from these distributions, the new (light blue) component displays characteristics of a \textit{thick disc} in the corresponding
maps in Figures~\ref{figure4} and ~\ref{figure4.1} (third row), while the new `disc' in dark blue looks like a \textit{thin disc}. 
The least rotationally supported component in this case (dark red) is more compact than the $nk=2$ spheroid, as it can be appreciated from the third rows of Figures~\ref{figure4} and ~\ref{figure4.1},
and as a consequence it is named \textit{bulge}. 

In the $nk=4$ case, third column of Figure~\ref{figure3} and second rows of Figures~\ref{figure4} and ~\ref{figure4.1}, 
parts of the least rotationally supported material of the $nk=3$ thick disc and of the $nk=3$ bulge are redistributed into a more extended velocity dispersion supported component,
namely a \textit{spheroid} (orange curves in Figure~\ref{figure3}). 
This spheroid basically has no net rotation and covers most of the binding energy range in an uniform manner. 

The largest value of $nk$ used throughout this study is $5$. For the test galaxy g8.26e11, $nk=5$ leads to an important redistribution of all the material in the components 
with large velocity dispersion support, including the thick disc. The one stable component is the \textit{thin disc}, whose definition changes the least from $nk=3$ to $nk=4$, and 
finally to $nk=5$. The very interesting aspect of this case is that now {\tt gsf} is able to separate the \textit{stellar halo} (magenta curves in the right column panels of Figure~\ref{figure3}), 
from the thick disc and the spheroid of $nk=4$. In the binding energy distribution, the {stellar halo} encompasses all the mass in the discreet low binding energy distribution peak 
and some material with $e/|e|_{\rm max}\geq-0.6$. These two different features in the binding energy distribution of the stellar halo correspond to the outer and inner components 
suggested by Milky Way observations \citep[e.g.][]{Carollo:2007}. From the circularity distributions, all the three components at low and negative $j_z/j_c$ have 
some degree of coherent rotation, see also the edge-on velocity maps in the top rows of Figures~\ref{figure4} and ~\ref{figure4.1}. 
Another interesting fact from $nk=5$ is the separation of the least from the most plane confined dispersion supported material, 
as it is evident from the dark red vs the red $j_p/j_c$ distributions in the center right panel of Figure~\ref{figure3}. 
Based on their appearance in the corresponding `observables' of Figures~\ref{figure4} and ~\ref{figure4.1}, 
the former is called \textit{classical bulge} and the latter \textit{pseudo bulge} \citep{Kormendy:2004}.  

To sum up, the {\tt gsf} run with $nk=5$ for the galaxy g8.26e11 results into a \textit{thin}, a \textit{thick disc}, a \textit{classical} and a \textit{pseudo bulge}, and a \textit{stellar halo}.
In the next sections we show that these kinematic structures agree to what theoretical models suggest as well as display properties seen in real data. 

\begin{figure*}
\begin{center}
\includegraphics[width=0.45\textwidth]{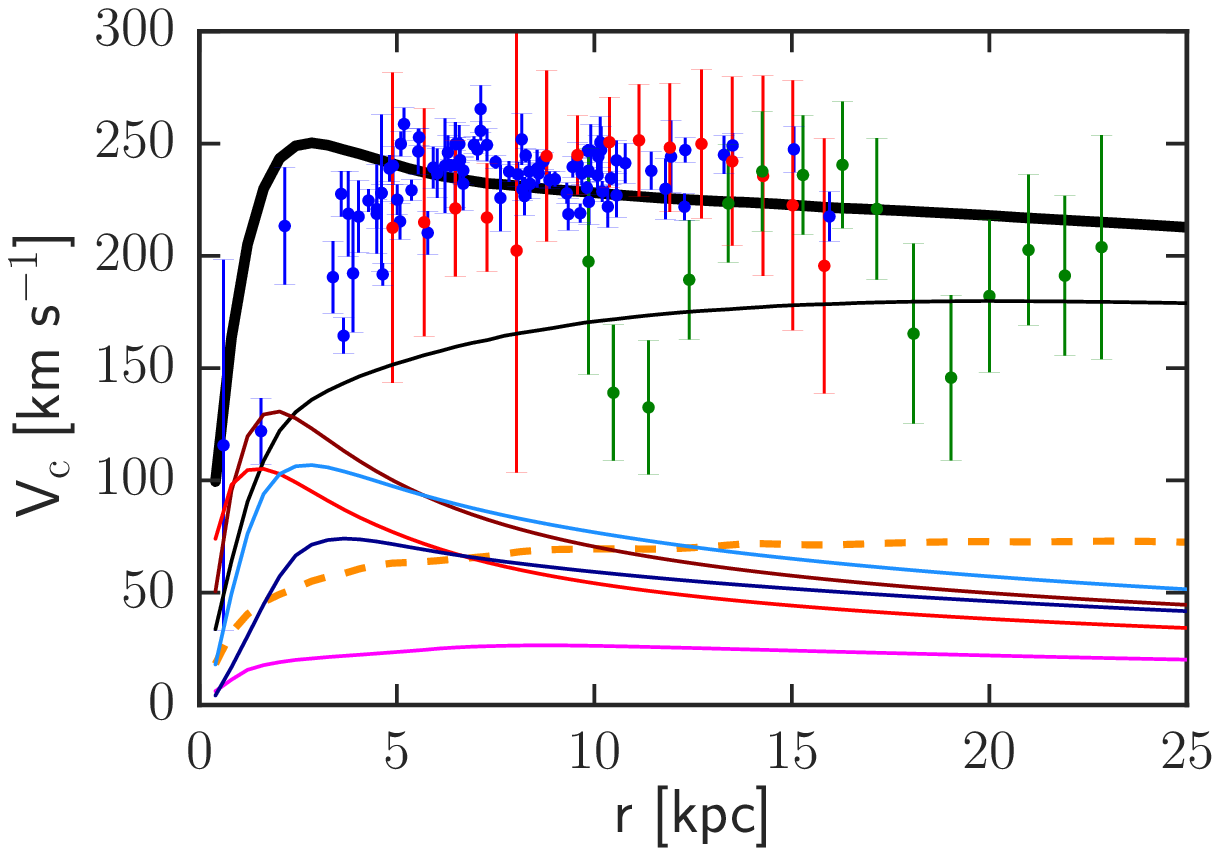}
\includegraphics[width=0.45\textwidth]{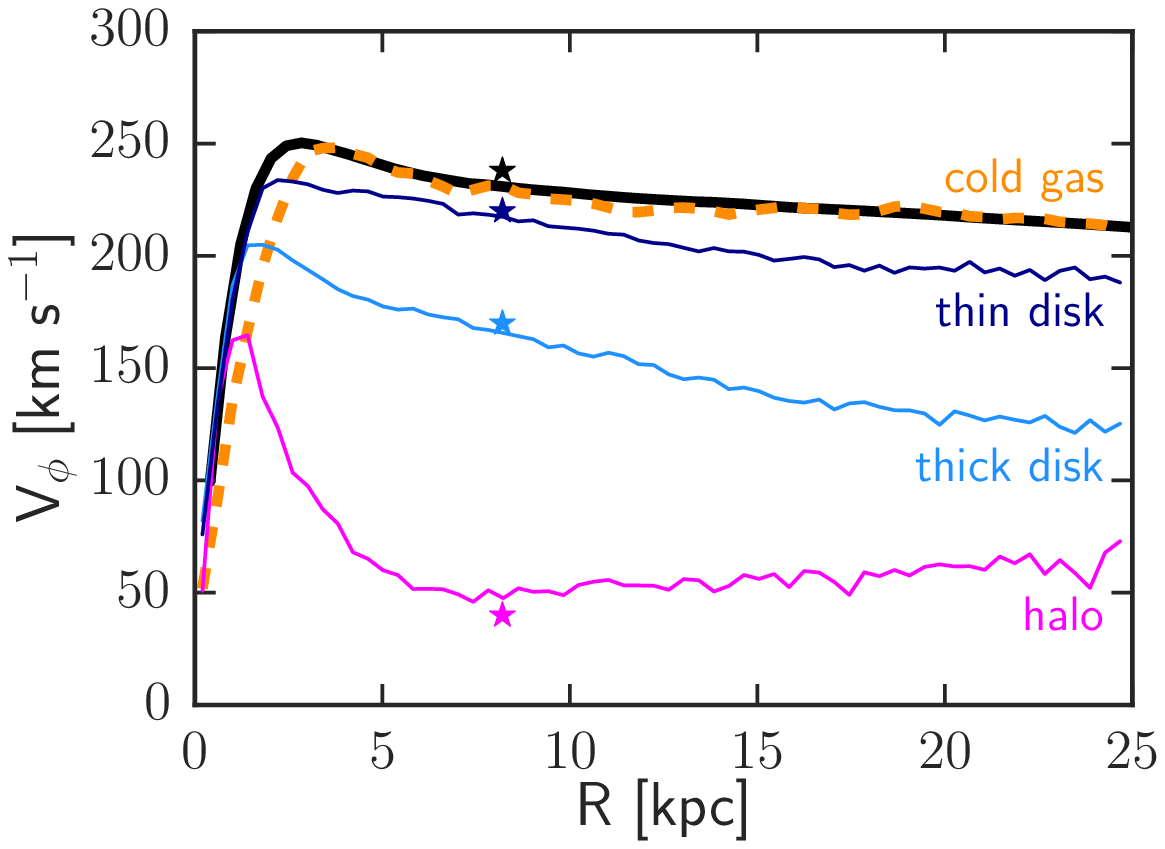}
\caption{\textbf{Left:} the contributions to g8.26e11's total circular velocity (thick black curve) of the various galaxy components (solid thin and dashed curves): 
dark matter (black), stellar halo (magenta), classical bulge (dark red), pseudo bulge (red), thick stellar disc (light blue), thin stellar disc (dark blue), 
and gas (dashed orange). The data points are observations of the Galaxy' stars by \citet{Reid:2014} (blue), \citet{LopezCorredoira:2014} (red) and \citet{Kafle:2012} (green), 
rescaled by \citet{Bland-Hawthorn:2016} to $238\pm15$~km~s$^{\rm -1}$ at $R_{\rm 0}=8.2$~kpc. 
At radii $r>5$~kpc, the total circular velocity of g8.26e11 is in very good agreement with the MW observations. 
\textbf{Right:} the radial profiles of the rotational velocities $V_{\rm\phi}$ for the thin (dark blue) and thick (light blue) stellar discs, the stellar halo (magenta) and the 
cold gas (dashed orange), and the total circular velocity of g8.26e11 (thick black curve). The coloured star symbols at $R=R_{\rm 0}$ give the MW observed rotational 
velocities in the solar neighbourhood of: the stellar halo ($\sim40$~km~s$^{\rm -1}$, magenta, \citealt{Bond:2010}), and of the thin and old thick stellar discs 
($220$~km~s$^{\rm -1}$ and $170\pm16$~km~s$^{\rm -1}$ in dark and light blue respectively, \citealt{Haywood:2013}). 
The black star symbol represents the circular velocity at the Sun's position ($238\pm15$~km~s$^{\rm -1}$, \citealt{Bland-Hawthorn:2016}, and references therein).}
\label{figure5}
\end{center}
\end{figure*}

\section{The solar neighbourhood perspective}
\label{z0prop_mw}

To have a quantitative assessment of how similar g8.26e11 and the Galaxy are, Figure~\ref{figure5} shows the total circular velocity profiles 
$V_c(r)=\sqrt{GM(<r)/r}$ and the contributions to it of the various stellar structures, gas and dark matter (left panel), and the profiles of the discs and stellar halo rotational velocities (right panel).
We use small $r$ to refer to the 3D radius, and capital $R$ for the projected one. 
Over plotted in the left panel are the MW total circular velocities derived from observations of: masers associated with massive young stars of \citet{Reid:2014} (blue points),
red clump giant stars of \citet{LopezCorredoira:2014} (red points), and blue horizontal branch star of \citet{Kafle:2012}. 
Given the fact that g8.26e11 was not simulated on purpose to resemble the MW, the agreement between its total circular velocity curve (thick black curve) 
and these observations is quite remarkable for radii $r>5$~kpc. 
At smaller radii, the MW has the Galactic bar, which is responsible for the dip in $V_c$ at $r\simeq3$~kpc
as probed by the dynamics of the HI gas \citep[e.g.][and references therein]{Sofue:2009}.
On the other hand the simulated galaxy has no bar, but a classical and a pseudo bulge, resulting in a purely rising $V_c$ at $r<2.5$~kpc.

In the right panel of Figure~\ref{figure5}, the total circular velocity curve $V_c$ of the simulated galaxy (thick black) is plotted together with the 
rotational velocities $V_{\phi}$ of the thin and thick stellar discs, stellar halo and cold gas. 
The rotational velocity $V_{\phi}$ profiles are computed as mass-weighted averages of $V$, 
where $V$ is the component of a particle's velocity along the direction of local rotation in the cylindrical coordinate reference frame of the galaxy's center.
We recall that the simulation is oriented with the $z$-axis in the direction of the total stellar angular momentum. 
The other two components of a particle's velocity in this reference frame are the radial $U$ and vertical $W$ velocities.  
One important feature obvious in this panel is that the total circular velocity is best traced by the cold gas rotation (dashed orange curve), 
the thin stellar disc (dark blue) having a $\sim20$~km~s$^{\rm -1}$ lower $V_{\phi}$. 
The dark blue and light blue stars on the plot represent measurements for the MW as published by \citet{Haywood:2013} for the solar neighbourhood thin and thick discs
of $220$~km~s$^{\rm -1}$ and $\sim170$~km~s$^{\rm -1}$, respectively, which are very close to the corresponding values of $V_{\phi}$ at the solar radius $R_0\simeq8.2$~kpc from the simulation
($218$ and $\sim166$~km~s$^{\rm -1}$, respectively). 
It is important to note that \citet{Haywood:2013} distinguish the two MW discs based on the stellar ages and positions in the [Fe/H]-[$\rm\alpha$/Fe] plane.
The magenta star gives the approximate stellar halo rotation at $R_0$ \citep{Bond:2010}, while the black star is the solar value of $V_c(R_0)\simeq238$~km~s$^{\rm -1}$ 
\citep[][and references therein]{Bland-Hawthorn:2016}.

One way to quantify a disc's thickness is through the vertical velocity dispersion. 
To compare the two discs of the simulated galaxies with the MW results, we selected a solar neighbourhood as $|R-R_0|<2$~kpc and $|z|<2$~kpc. 
The velocity dispersion in the vertical direction $\sigma_W$ of the g8.26e11's thick disc at $R_0$ is $73$~km~s$^{\rm -1}$ and of the thin disc is $29$~km~s$^{\rm -1}$. 
The various studies of the MW did not yet converged on one value for each of the two discs' $\sigma_W$ given the differences in the surveys selection functions, 
sky coverages, dynamical modeling, and thin/thick definition. 
For the MW thick disc, \citet{Robin:2017} found values as low as $27$~km~s$^{\rm -1}$, while \citet{Binney:2012} obtained $\sigma_W$ in the range [31,65]~km~s$^{\rm -1}$. 
\citet{Robin:2017} also found the lowest values for the MW thin disc, between $6$ and $20$~km~s$^{\rm -1}$, while \citet{Binney:2012} found values in the range [20,27]~km~s$^{\rm -1}$. 
We can therefore conclude that g8.26e11 is a realistic MW analogue from the point of view of the solar neighbourhood stellar population dynamics.

Globally the stellar mass of g8.26e11 is distributed as follows: $21\%$ in the thin disc, $33\%$ in the thick disc, 
$25\%$ in the classical bulge, $14.5\%$ in the pseudo bulge, and the remaining $6.5\%$ in the stellar halo.
Therefore, from a dynamics point of view, g8.26e11 has a bulge-to-total ($B/T$) mass ratio of 0.46, summing up the contributions of the two bulges and the stellar halo. 
For comparison, the Milky Way is estimated to have $B/T\rm\simeq0.30$ \citep[see review by][]{Bland-Hawthorn:2016}. 

\begin{figure*}
\begin{center}
\includegraphics[width=0.45\textwidth]{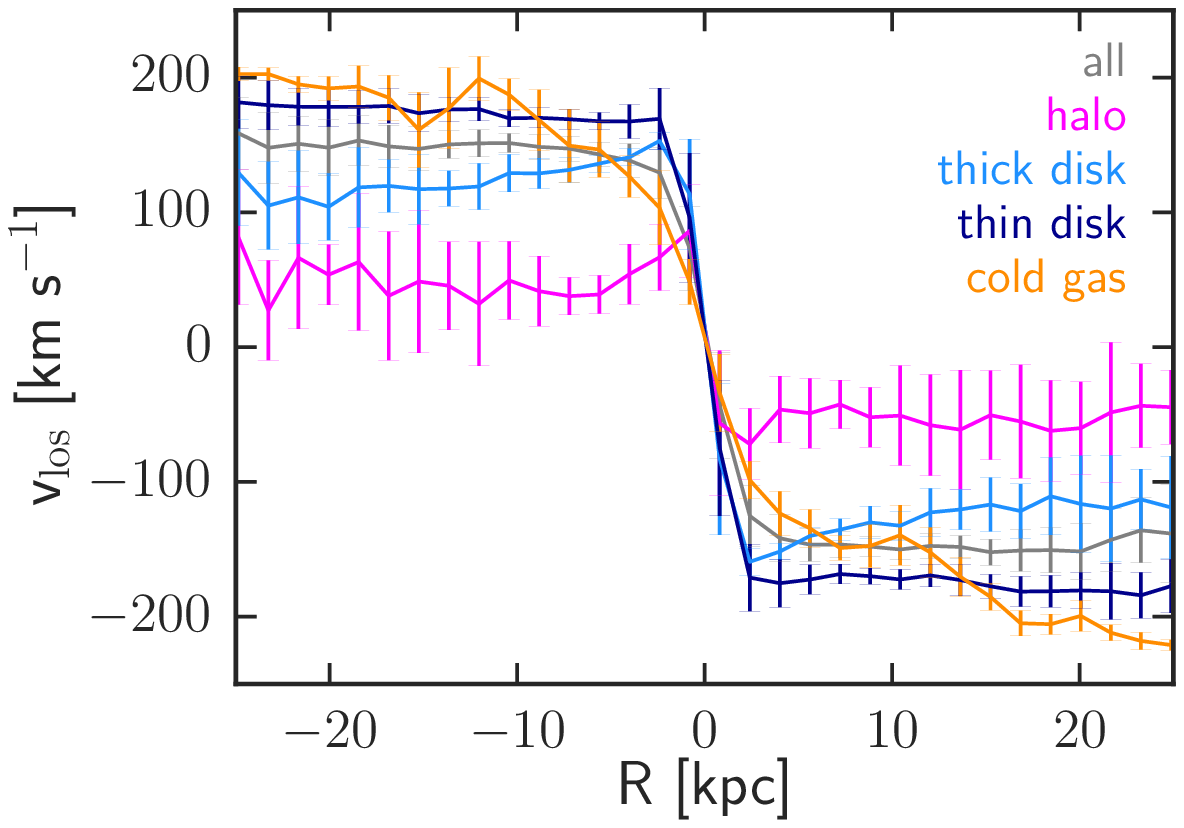}
\includegraphics[width=0.45\textwidth]{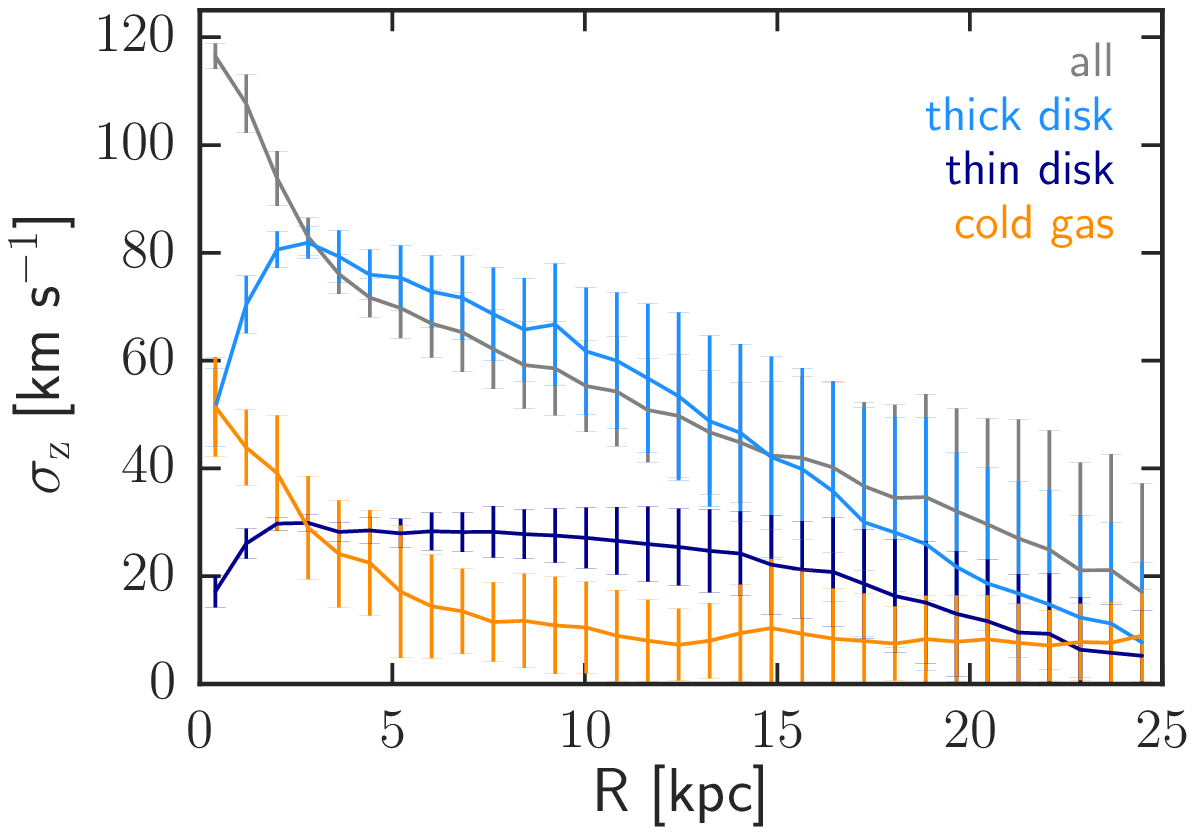}
\caption{The radial profiles of line-of-sight velocities $v_{\rm los}$ assuming an edge-on perspective (left) and vertical velocity dispersions $\sigma_{\rm z}$ (right) 
for the components of galaxy g8.26e11. The stellar components are shown with the same colors as in the right column of Figure~\ref{figure3}, while the cold gas is given in orange.}
\label{figure6}
\end{center}
\end{figure*}

\begin{figure}
\begin{center}
\includegraphics[width=0.45\textwidth]{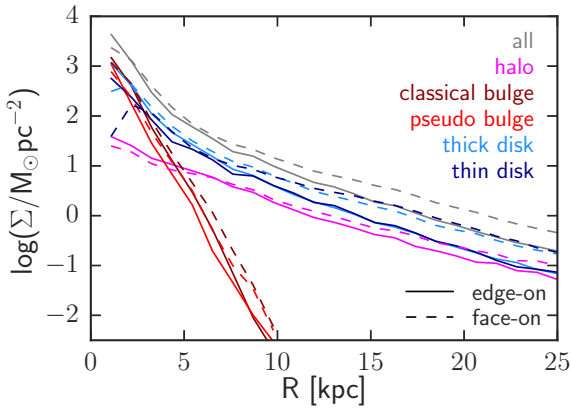}
\caption{The radial profiles of surface mass densities $\Sigma$ for the components of galaxy g8.26e11. The stellar components are shown with the same colors 
as in the right column of Figure~\ref{figure3}. The solid and dashed lines in the right panel give the edge-on and face-on on perspectives, respectively.}
\label{figure6.1}
\end{center}
\end{figure}

\section{The simulated galaxy seen as an extragalactic object}
\label{z0prop}

In external galaxies, however, the profiles of $V_{\phi}$, as shown in the right panel of Figure~\ref{figure5}, can not be directly measured. 
For extragalactic objects the kinematic information comes instead in the form of line-of-sight velocity $v_{\rm los}$ fields like the one shown in the bottom row of Figure~\ref{figure4.1}
for the total stellar population of g8.26e11. 

In order to compare the simulation with observed external galaxies, radial profiles of line-of-sight velocity $v_{\rm los}$ were extracted   
along the major axis (horizontal) in Figure~\ref{figure4.1}, using a slit of 1.6~kpc width. 
The left panel of Figure~\ref{figure6} shows the radial profiles of $v_{\rm los}$ for the thin (dark blue) and thick (light blue) discs, stellar halo (magenta), 
all stars (grey) and cold gas (orange) of the simulated galaxy g8.26e11. 
One of the first thing to notice in this panel is that the $v_{\rm los}$ profiles for both thin disc (dark blue) and all the stars (grey) are flat after $R\gtrsim5$~kpc,
the former saturating at $174\pm5$~km~s$^{\rm -1}$ and the latter at $149\pm4$~km~s$^{\rm -1}$. 
A similar trend can be observed for the stellar halo which saturates at $v_{\rm los}=44\pm15$~km~s$^{\rm -1}$
On the contrary, the thick disc has a line-of-sight velocity profile declining with the radius, on average $\sim50$~km~s$^{\rm -1}$ lower than that of the thin disc.   
In observed external galaxies the $v_{\rm los}$ of the stellar haloes can not be measured at such small radii ($R<25$~kpc) given that the convolved velocity is heavily dominated by the disc(s),
and that stellar haloes are much fainter than the central components.

We also note that the $v_{\rm los}$ profiles for the thin and thick discs are significantly below their corresponding $V_{\phi}$ at all radii $R$. The same is true for the cold gas.
These differences in rotational velocities can have important consequences on what is inferred from observational studies on the intrinsic stellar distribution, 
and consequently on the inner dark matter halo mass estimates.
What is typically accessible in observations of extragalactic stellar velocity fields are absorption lines produced mainly in the atmospheres of young stars 
\citep[e.g.][]{Yoachim:2008b,Martinsson:2013}.
However, if the rotational velocity profiles  derived in this manner are assumed to be a characteristic of the full stellar distribution, 
the galaxies will be taken to be more dynamically cold, i.e. discy, than they truly are. 
Therefore, we think that circular velocities of external galaxies derived from gas and/or stellar kinematics might also be significantly underestimated. 
Quantifying this bias is, however, beyond the scope of this paper.

The right panel of Figure~\ref{figure6} shows another galaxy observable, the vertical velocity dispersion $\sigma_z$ profile. The thin and thick discs, all stars and cold gas 
vertical velocity dispersions are given by the dark blue, light blue, grey and orange curves, respectively. 
As expected of thin stellar discs, the vertical velocity dispersion $\sigma_z$ profile decreases slowly with radius, and can be approximated as a constant of $\sim25$~km~s$^{\rm -1}$. 
The thin disc $\sigma_z(R)$ looks very different than the one of the whole galaxy, which is much better approximated by the thick disc at $R\geq5$~kpc. 
The thick disc $\sigma_z(R)$ decreases approximately linearly with the projected radius, from $\sim75$~km~s$^{\rm -1}$ at $R\sim5$~kpc to $\sim20$~km~s$^{\rm -1}$ at $R\sim20$~kpc. 
The central peak of the whole galaxy's $\sigma_z(R)$ is produced by the bulge components and reaches $\sim120$~km~$s^{\rm -1}$ in the very center. 
The cold gas' $\sigma_z$ decreases approximately linearly between $\sim50$~km~s$^{\rm -1}$ in the center to $\sim10$~km~s$^{\rm -1}$ at $R=5$~kpc, after which it stays constant. 
Overall, from both normalization and the shape of the edge-on line-of-sight velocity profiles and vertical velocity dispersion profiles, g8.26e11 resembles the observed galaxy UGC 00448 
after correcting it for inclination effects \citep[Appendix D of][]{Martinsson:2013}. UGC 00448 is a less massive galaxy than the simulated one, 
with a stellar mass of $1.9\pm1.0\times10^{\rm 10}$M$_{\rm\odot}$ derived using the HI line width of \citet{Staveley-Smith:1988} and the Tully-Fisher relation of \citet{Dutton:2017}.

Figure~\ref{figure6.1} gives the total stellar surface mass density $\Sigma$ profile (grey curves) in 
both edge-on (solid grey) and face-on (dashed grey) perspectives. To recover this type of profile from photometry of observed galaxies, one has to assume 
a mass-to-light ($M/L$) ratio(s). The same panel also shows the corresponding profiles in both perspectives for all five stellar kinematic components. 
The total stellar surface mass density is peaked in the centre ($R<5$~kpc) and well fitted by an exponential at larger radii ($5<R<25$~kpc), 
with scalelengths $R_{\rm d}$ of $3.5\pm0.1$ and $4.1\pm0.1$~kpc in edge-on and face-on perspectives, respectively. 
For comparison, the scalelength of UGC 00448 in the K-band face-on corrected profile is $3.9\pm0.2$~kpc \citep{Martinsson:2013}.
UGC 00448 shows a very similar radial light distribution to g8.26e11 stellar mass one, with a central peak and a purely exponential profile for $R>4.3$~kpc. 
This observed galaxy is classified as SABc. 

Looking at the contributions of the five components to the $\Sigma(R)$ in edge-on perspective, the two discs extend all the way to $R=0$, but do not have purely exponential profiles. 
The disc peak in the centre is a behavior expected from the evolution of thin discs that conserve their angular while reaching for their minimum energy state \citep{Lynden-Bell:1972}.
In face-on perspective the two disc types have a dip in the centre. 
These dips are a direct consequence of the GMM algorithm which associates very small probabilities for particles in the very inner region to belong to the discs.
Fitting with an exponential the profiles for the two discs in the range $5<R<25$~kpc, we obtained scalelengths $R_{\rm d}$ of $3.5\pm0.1$ and $3.4\pm0.1$~kpc 
for the thin and thick discs in edge-on perspective, and $4.3\pm0.1$ and $3.9\pm0.1$~kpc in face-on perspective, respectively. 
In the case of the Galaxy, the two discs have $R_{\rm d}$ in a similar range, $R_{\rm d}=2.7-3.7$~kpc \citep[e.g.][]{Piffl:2014,Sanders:2015,Binney:2015}.

Interestingly, the three components supported by random motions in both face-on and edge-on perspective, 
namely the classical and pseudo bulges and the stellar halo, show purely exponential profiles. 
Though this is expected of structures like low mass spheroidal galaxies \citep{Graham:2003,Koda:2015,vanDokkum:2015} or pseudo bulges \citep{Fisher:2008,Gadotti:2009}, 
it is not generally expected of classical bulges, which are thought to have S\'{ersic} indices $n_{\rm S}\geq2$. 
However, other authors like for example \citet{Andredakis:1994} or \citet{Andredakis:1995} 
have argued that bulges cover a continuous range in profiles from purely exponential to highly centrally concentrated ones.

If one would try to fit $\Sigma(R)$ of all stars, a combination of an exponential and a S\'{ersic} profiles, or even a S\'{ersic} profile only would suffice.
The observational bias towards large $n_{\rm S}$ for bulges can be partially explained by the expectation that the disc is purely exponential and reaches the center. 
This requirement for the disc forces the central component to a concave fitting function, i.e. a large $n_{\rm S}$. 
However, the parameters resulting from such fits can severely bias the conclusions drawn on galaxy formation \citep[e.g.][]{Mosenkov:2014,Bernardi:2014}.

\begin{figure}
\begin{center}
\includegraphics[width=0.45\textwidth]{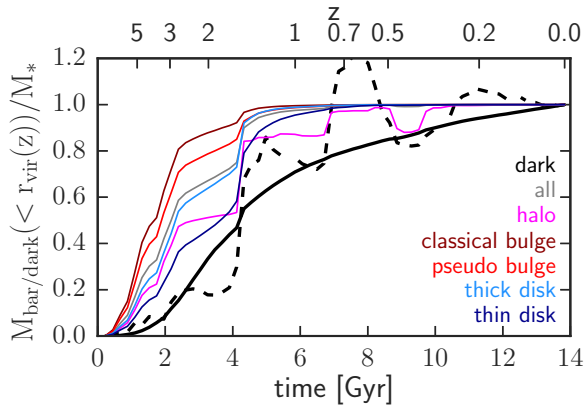}
\caption{The assembly of baryons and dark matter along the main branch of the merger tree. 
The thick black curve and the coloured solid ones show the evolution of the dark matter halo mass normalized to its value at $z=0$ and 
of the normalized progenitor baryonic masses inside the viral radius along the main branch of the merger tree for the $z=0$ components of g8.26e11.
The thick dashed black curve represents the evolution of the dark matter halo specific angular momentum $j_{\rm h}$ normalized to its final value.}
\label{figure8.1}
\end{center}
\end{figure}

\begin{figure*}
\begin{center}
\includegraphics[width=0.45\textwidth]{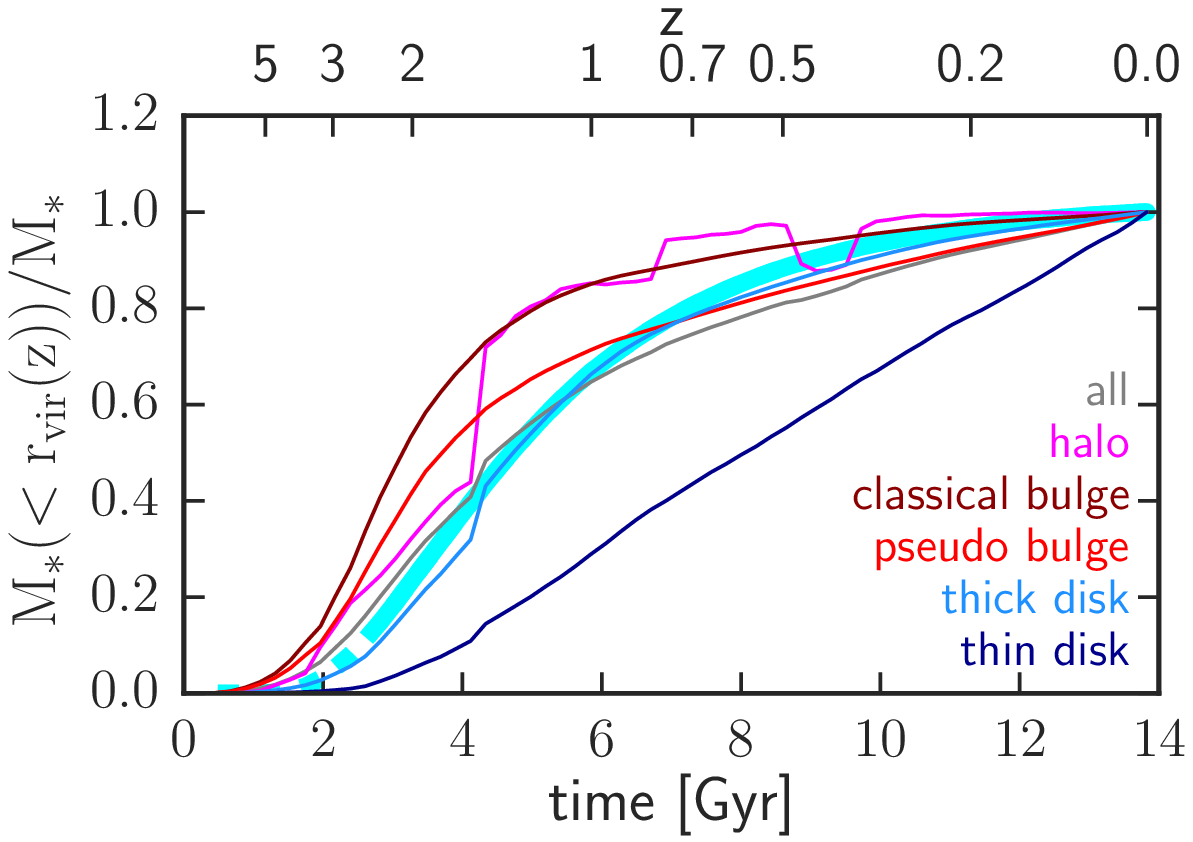}
\includegraphics[width=0.45\textwidth]{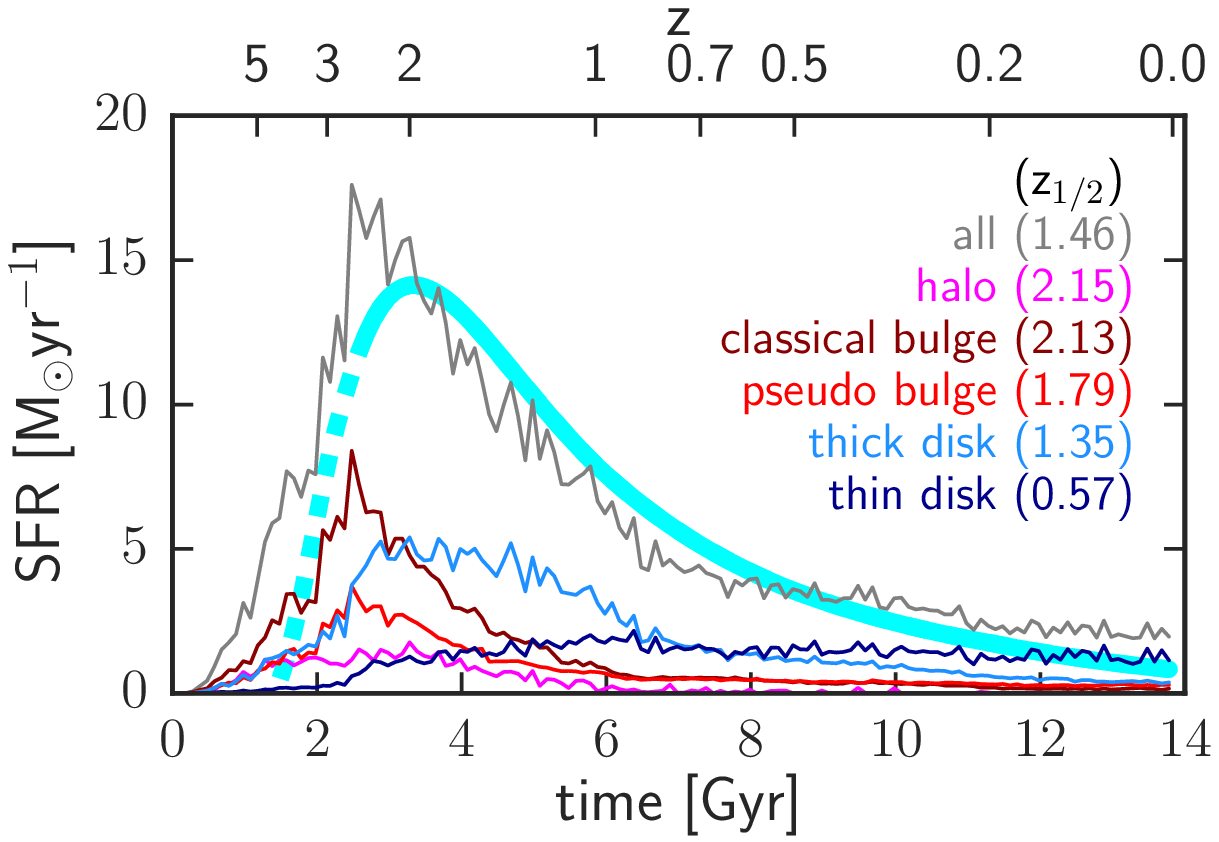}
\caption{The evolution of the normalized stellar masses inside the viral radius along the main branch of the merger tree (left) and the star formation rate histories for the components of g8.26e11.
The normalizations of each component (left) is done with respect to its corresponding stellar mass at $z=0$.
The thick cyan curves are the corresponding parameterizations of \citet{vanDokkum:2013} for a MW mass galaxy, 
while the thick dashed cyan ones are their extrapolations to higher redshifts. 
The colored numbers after each name in the right panel give the half stellar mass formation redshifts $z_{\rm 1/2}$.}
\label{figure8}
\end{center}
\end{figure*}

\section{Formation of stellar structures}
\label{evolution}

In order to provide a link with the dark matter halo evolution, 
Figure~\ref{figure8.1} shows the dark matter mass (thick solid black) and specific angular momentum (thick dashed black) build up along the main branch of the merger tree. 
Both quantities are normalized to their respective values at $z=0$. 
The figure also gives the normalized progenitor baryonic masses inside the viral radius along the main branch of the merger tree for the $z=0$ components of g8.26e11 (thin coloured curves).
The baryonic mass assemblies look like step functions, 
while the dark matter halo mass growth is very smooth, with only one small jump at $z\sim1.6$, which marks the transition 
from a large growth rate to an ever decreasing small one. This small step represents the last important merger at the dark matter halo scale. 
The major merger at $z\sim1.6$ brings in almost half of the final baryonic mass of the thin disc, but less than $10\%$ of the classical bulge mass. 
Also, from the thick dashed black curve, this merger is responsible for a large fraction of the dark matter halo spin, $\geq60\%$. 

Figure~\ref{figure8.1} supports the idea that the kinematic structures themselves also get assembled in the same temporal sequence as their mass gets transformed into stars. 
Basically, the classical bulge forms first, followed by the pseudo bulge, then by the thick disc and later by the thin disc. 
In this perspective, the stellar halo does not fit into this sequence, showing a much more syncopated assembly pattern. 
Counting the large jumps in the baryonic mass assembly of the stellar halo (magenta curve) one can say that almost $80\%$ of its progenitor material came in with 
the minor mergers at $z\sim3.5$ and $0.8$ and with the major merger at $z\sim1.6$. 
This behaviour together with the significant jumps in the its corresponding normalized stellar mass (magenta curve) is a clear indication that a large fraction of this 
$\sim80\%$ stellar halo progenitor mass came inside $r_{\rm vir}$ as already formed stars.

Figure~\ref{figure8} shows the star formation rate, hereafter SFR, (right)  
and the normalized stellar masses inside the viral radius along the main branch of the merger tree (left) for the $z=0$ components of g8.26e11. 
The SFR of the whole galaxy in grey shows a prominent peak between $z\sim3$ and $z\sim2$, 
a fast decrease in intensity between $z\sim2.5$ and $z\sim0.9$, 
and a much slower decrease afterwards down to a $z=0$ value of $\sim2M_{\rm\odot}yr^{\rm -1}$.
Over plotted in cyan is the reconstructed SFR history for a MW mass galaxy of \citet{vanDokkum:2013}. 
This observationally derived SFR history resembles well both the shape and the normalization of g8.26e11 (grey curve), apart from a slight shift of the simulation's SFR towards earlier times.

Looking at the five components separately, the three dispersion supported ones show clear high redshift, $z>2$, SFR peaks, 
and very little (the two bulges) to no SFR (the stellar halo) after $z\sim1$. 
The thin disc grows its stellar mass at an approximately constant rate of $1.5M_{\rm\odot}yr^{\rm -1}$ only after the peaks of the bulges and of the stellar halo. 
The thick disc, on the other hand shows more similarities with the dispersion dominated components than with the thin disc, 
although it reaches its maximum SFR later on, and the decrease at lower redshifts is more gradual. 
The fact that the thick disc forms most of its stars at early times is in agreement with the scenario proposed by \citet{Brook:2004}, 
who found that the disc stars with higher vertical velocities tend to be preferentially born during the high redshift epoch of frequent, chaotic mergers.  

The formation times of the stars in the five kinematic structures of g8.26e11 show a clear trend. 
The colored numbers in the upper right corner of the SFR panel of Figure~\ref{figure8} give the components' half stellar mass formation redshifts $z_{\rm 1/2}$. 
These values tell the story of a stellar mass formation sequence, with the stars of the halo forming first ($z_{\rm 1/2}=2.15$), 
followed by those of the classical bulge ($z_{\rm 1/2}=2.13$), the pseudo bulge ($z_{\rm 1/2}=1.79$), the thick disc ($z_{\rm 1/2}=1.35$) 
and finally those of the thin disc ($z_{\rm 1/2}=0.57$).
The $z_{\rm 1/2}=1.46$ of the whole galaxy is in between the pseudo bulge and the thick disc values. 
This sequence, however, does not necessarily imply that the kinematic structures themselves formed, understood as in `got assembled', in this order. 

The left panel of Figure~\ref{figure8} represents the stellar mass assembly for the various components of g8.26e11. 
The thin disc and the two bulges show smoothly increasing curves, suggesting that the \textit{SFR occurred in-situ} for these three stellar structures. 
Interestingly, this formation pattern for classical bulges disfavours the merger scenario \citep[e.g.][]{Aguerri:2001} in the particular case of this simulated galaxy.
The thick disc, however, shows a relatively large jump at the same redshift $z\sim1.6$ of the largest jump visible for the stellar halo.
These feature are a clear indication of a merger, which results in the stars of the infalling object to be incorporated later on to either the thick disc or the stellar halo of the main galaxy.
Globally, the stellar halo has the largest fraction of stars \textit{born ex-situ} ($45\%$), followed by the thick disc with $8\%$, and finishing with the thin disc with only $2\%$.
Both classical and pseudo bulges of this galaxy formed all their stars in-situ.

Same as for the SFR, we over plotted in the left panel of Figure~\ref{figure8} the observationally constrained stellar mass assembly for a MW mass galaxy of 
\citet{vanDokkum:2013} (cyan curve). This observational curve is relatively close to the grey curve representing the global evolution of the simulated galaxy. 
At closer inspection, the component most similar to the observations is the thick disc. 
Therefore, g8.26e11 not only resembles the MW at redshift $z=0$, but also has an assembly/SFR history history very similar to what the observational study of \citet{vanDokkum:2013} suggests.
This plot shows clearly a stellar mass assembly sequence very similar to the SFR history one, excluding the stellar halo. 
The stellar halo of g8.26e11 is to a great extent a product of mergers.

\subsection{Evolution of spins, sizes, shapes and rotational support}
\label{prop_ev}

The properties of the kinematic structures of g8.26e11 described before refer to the galaxy at redshift $z=0$. 
For the study of the various properties' evolutions we trace back the Lagrangian mass defined at $z=0$ as belonging to either one of the kinematic components, or to the galaxy (stars) as a whole. 
The aim is to quantify how the five kinematic components of g8.26e11 evolve in the properties that discriminate among them, 
namely: sizes, shapes, angular momenta and rotational support. All these quantities are computed in physical units. 

For each component $k$ at a given time $t$, we first calculate its center of mass position $\vec{\rm r}_{\rm k}{\rm(t)}$, and velocity $\vec{\rm v}_{\rm k}{\rm(t)}$, using 
all the baryon particles $\{i\}$ that are progenitors of the stellar particles of component $k$ at $z=0$, $\{i\}(t)\rm\Leftrightarrow k(z=0)$, 
and update the positions $\vec{\rm r}_{\rm i}{\rm(t)}$, and velocities $\vec{\rm v}_{\rm i}{\rm(t)}$ with respect to center of mass reference frame.
The masses $m_{\rm i}(t)$ used for the particles $\{i\}$ at any time, are their corresponding stellar masses from $z=0$, $m_{\rm i}(t) = m_{\rm i(*)}(z=0)$.
In order to have a more straightforward interpretation of the dynamical quantities' evolutions the orientation of the simulation box at each time step is kept fixed, 
such that the $z$-axis is parallel to the total stellar angular momentum of the galaxy at redshift $z=0$. 

We use the term \textit{size} to refer to the 3D half mass radius $r_{\rm 50}(k;t)$, and the term \textit{shape} for 
the ellipticity defined as: 
\begin{equation}
\varepsilon(k;t)=1-\frac{c(k;t)}{a(k;t)},  
\end{equation}
where the semiaxes $a$ and $c$ are computed from the eigenvalues $E_1\leq E_2\leq E_3$ of the inertia tensor $I_{\rm jl}(k;t)$ of the particles $\{i\}(t)$, 
following \citet{GonzalezGarcia:2005}. The inertia tensor is defined as:
\begin{equation}
I_{jl}^{(k)} = \sum_{i\in(k)} m_i(\delta_{jl}r_i^2-x_jx_l),
\end{equation}
with $j$ and $l$ looping over the Cartesian coordinates. The semiaxes $a>b>c$ are computed from:
\begin{equation}
  \label{eq7}
  \begin{aligned}
    a^2+b^2+c^2 &= 5(E_1+E_2+E_3)/2\\
    a^2/b^2 &= (E_3+E_2-E_1)/(E_1+E_3-E_2)\\
    a^2/c^2 &= (E_3+E_2-E_1)/(E_1+E_2-E_3)
  \end{aligned}
\end{equation}

The specific angular momentum $j(k;t)$ will be referred to as \textit{spin} of the component $k$, given that it is computed with respect to the reference frame of the particle group $\{i\}(t)$:
\begin{equation}
j(k;t) = \frac{| \sum_{i\in(k)} m_i\vec{r}_i(t)\times \vec{v}_i(t)|}{ \sum_{i\in(k)} m_i},
\end{equation}

To estimate the amount of rotational support we use the velocity dispersion fraction $f_{\sigma}(k;t)$ defined as:
\begin{equation}
f_{\sigma}(k;t)=1-3\frac{\sigma_z(k;t)^2}{\sigma(k;t)^2}. 
\end{equation}
For a group of particles with isotropic velocities in their own reference frame, we expect this fraction to be zero because $\sigma_z\simeq\sigma/\sqrt{3}$.
If the velocities of the particles are instead confined to the $xy$-plane, $\sigma_z\simeq0$, and the fraction should be one. 
We use $f_{\sigma}$ to estimate the rotational support as opposed to the more conventional $v/\sigma$ \citep{Davies:1983} because the latter can not be employed at high redshift, 
where the Lagrangian masses of the components extend way outside the virial radius of the progenitor halo.  

\begin{figure}
\begin{center}
\includegraphics[width=0.45\textwidth]{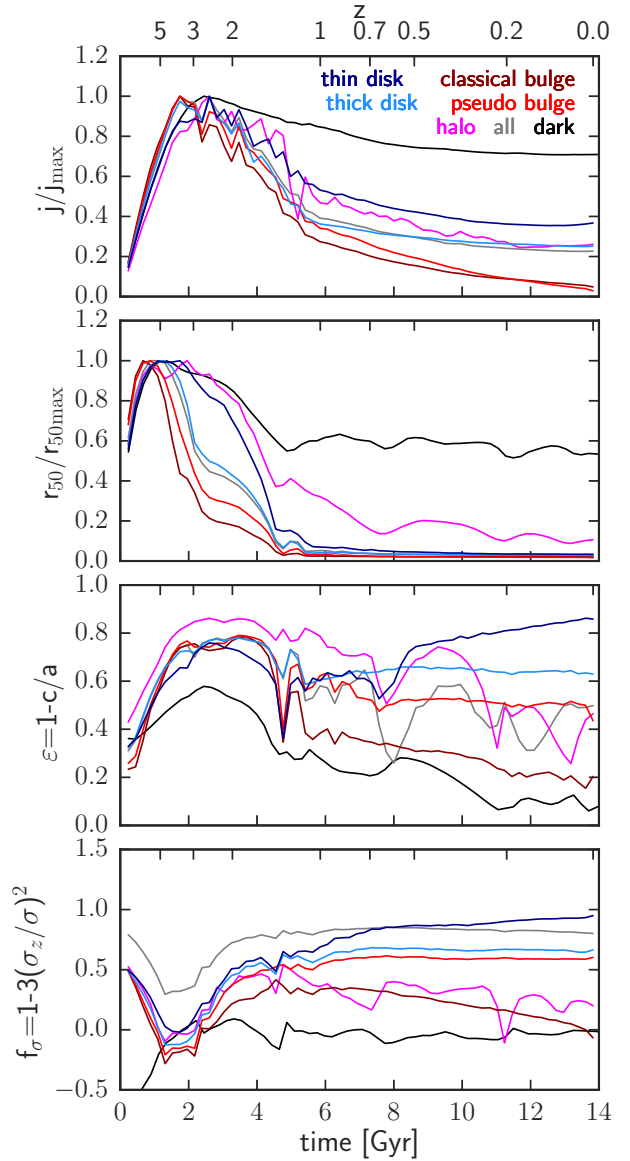}
\caption{Evolutions of the normalized spins (top), normalized sizes (centre top),
shapes (centre bottom) and vertical velocity dispersion fractions 
(bottom) for the various components of g8.26e11.
The quantities plotted have been calculated by tracing back the Lagrangian mass of each kinematic group separately. 
The black lines in all panels show the corresponding evolution of the same property for the $z=0$ dark matter halo Lagrangian mass.}
\label{figure7}
\end{center}
\end{figure}

The colored curves of Figure~\ref{figure7} show the evolutions of the spins (top), sizes (centre top), shapes (centre bottom) and velocity dispersion fractions (bottom)
for the Lagrangian masses of each of the five kinematic components of g8.26e11. 
The grey curves give the corresponding evolutions of all the stellar particles of the galaxy at $z=0$, while the black ones represent the Lagrangian mass of the $z=0$ dark matter halo. 
Both spins and sizes have been normalized to their respective maximum values to ease the comparison between the various galaxy components. 

At early epochs the spins of all components grow approximately linearly with time until they reach their maximum values, at redshifts around $3$.
This early behavior reproduces well the predictions of the tidal torque theory \citep{Hoyle:1951,Peebles:1969,Doroshkevich:1970,White:1984}, 
which links the angular momentum acquisition in protogalaxies with the torques induced upon each other by neighbouring collapsing regions of the universe. 
In this framework, a collapsing region is expected to attain its maximum angular momentum when it reaches its maximum extent and its 
evolution decouples from the universal expansion. In the spherical collapse model, this time is called turn-around. 
Therefore, we can identify the beginning of the g8.26e11 halo collapse with this turn-around redshift, $z_{\rm turn}\sim3$. 
After this time, all components lose part of their angular momenta, with the dark matter losing only $\sim30\%$, while the two bulges lose more than $95\%$. 
Among the five kinematic components, the thin disc loses the least, $\sim60\%$. 
A qualitatively similar specific angular momentum evolution has been shown by \citet{Dominguez:2015} for the dynamical disc and spheroid components of two simulated galaxies.

As we already showed in Section~\ref{z0prop_mw}, the total circular velocity $V_{\rm c}$ of the simulated galaxy is very similar to that of the MW.  
Also its rotational velocities $V_{\phi}$ for its various stellar components are in very good agreement with MW observations in the solar neighbourhood, 
while the disc(s) scalelengths are in agreement with both MW and external galaxy measurements.
In the light of these results, the loss of angular momentum we found is a genuine property of the baryonic collapse and assembly, 
and not an effect of the so-called `angular momentum problem' \citep{Navarro:2000}, which affected earlier generations of simulations. 
This problem has largely been solved by improving the numerical schemes \citep[e.g][]{Serna:2003}, 
increasing the resolution \citep[e.g][]{Governato:2004} and implementing feedback processes \citep[e.g.][]{Okamoto:2005}.

The merger at $z\sim1.6$ is easy to identify in both the sizes evolution plot (centre top of Figure~\ref{figure7}) and the shapes one (center bottom). 
This epoch marks the halo virialization, $z_{\rm vir}\sim1.3$, as exemplified by the dark matter $r_{\rm 50}$ reaching its equilibrium value, 
and by the sharp dips in the evolution of the thin disc's and classical and pseudo bulges' shapes, $\varepsilon$. 
The collapse of the various kinematic components follows the same sequence as their SFR histories, with the classical bulges being first and the thin disc last.
Same as before, the stellar halo does not follow the evolution of the other kinematic components, instead resembling more the dark matter, 
i.e. the wiggles in the size evolutions of the stellar halo and the dark matter halo are correlated. 

All five kinematic components of g8.26e11 lose angular momentum faster between $z_{\rm turn}\sim3$ and $z_{\rm vir}\sim1.3$ than later on. 
With the exception of the thin disc, the other four also form a large fraction of their stars during the same epoch. 
These stars are thus formed in a highly turbulent environment, i.e. the collapsing dark matter halo(es), where the gravitational potential varies on short timescales. 
This suggests that the dominant physical process responsible for the stellar loss of angular momentum during this time is violent relaxation \citep{LyndenBell:1967}. 
Asymmetries of the gravitational potential generated by mergers also provide an efficient mechanism to transfer the gas angular momentum. 
Gas can also lose angular momentum through dynamical friction \citep{Chandrasekhar:1943,Leeuwin:1997}, 
a physical process whose effects in simulations depend both on the resolution and on the numerical algorithms employed \citep{Semelin:2002}.  

One possible explanation for the little loss of angular momentum by the thin disc component 
is its progenitor gas being part of the hot halo previous to its arrival on the equatorial plane \citep{Athanassoula:2016}. 
\citet{Peschken:2017} find in prepared merger simulations that this seems to be the case, the angular momentum of the discs increasing with time 
at the expense of the angular momentum of the gaseous halo \citep{Eggen:1962}.
While the progenitor gas of the thin kinematic disc of g8.26e11 might have passed through the hot halo phase, our results suggest that the main cause for its angular momentum conservation 
is simply the fact that a large part of this material accretes onto the galaxies at times when the dark matter halo 
is already virialized, and as such there is no physical mechanism that is able to alter it considerably.

In the bottom two panels of Figure~\ref{figure7} we estimate how much disc-like in shape ($\varepsilon$ close to $1$) and in rotational support 
($f_{\sigma}$ close to 1) the five components, the whole galaxy and the dark matter halo are.  
At high redshifts all components have high values of $\varepsilon$ because their material is part of the filamentary large scale structure. 
As the dark matter gets assembled, its shape evolves towards spherical symmetry ($\varepsilon\sim0$) and its rotational support settles to zero, as expected. 
For all baryonic components shown, from $z_{\rm turn}\sim3$ to $z_{\rm vir}\sim1.3$ the shapes evolve towards spherical symmetry, while the rotational support increase.
After $z_{\rm vir}$, different behaviors emerge. 
The material of the classical bulge loses all its rotational support, ending up as a velocity dispersion dominated system with a small ellipticity $\varepsilon\sim0.2$.
The pseudo bulge and the thick disc show almost no evolution between $z_{\rm vir}\sim1.3$ and $z=0$, the final ellipticity of the former being $\sim0.45$ and of the latter $\sim0.65$.
The thin disc on the other hand increases its $\varepsilon$ up to $\sim0.85$ and its $f_{\sigma}$ up to one. 
From these two physical properties at $z=0$ it clearly qualifies for the nickname of `thin disc'. 

\begin{figure*}
\includegraphics[width=0.98\textwidth]{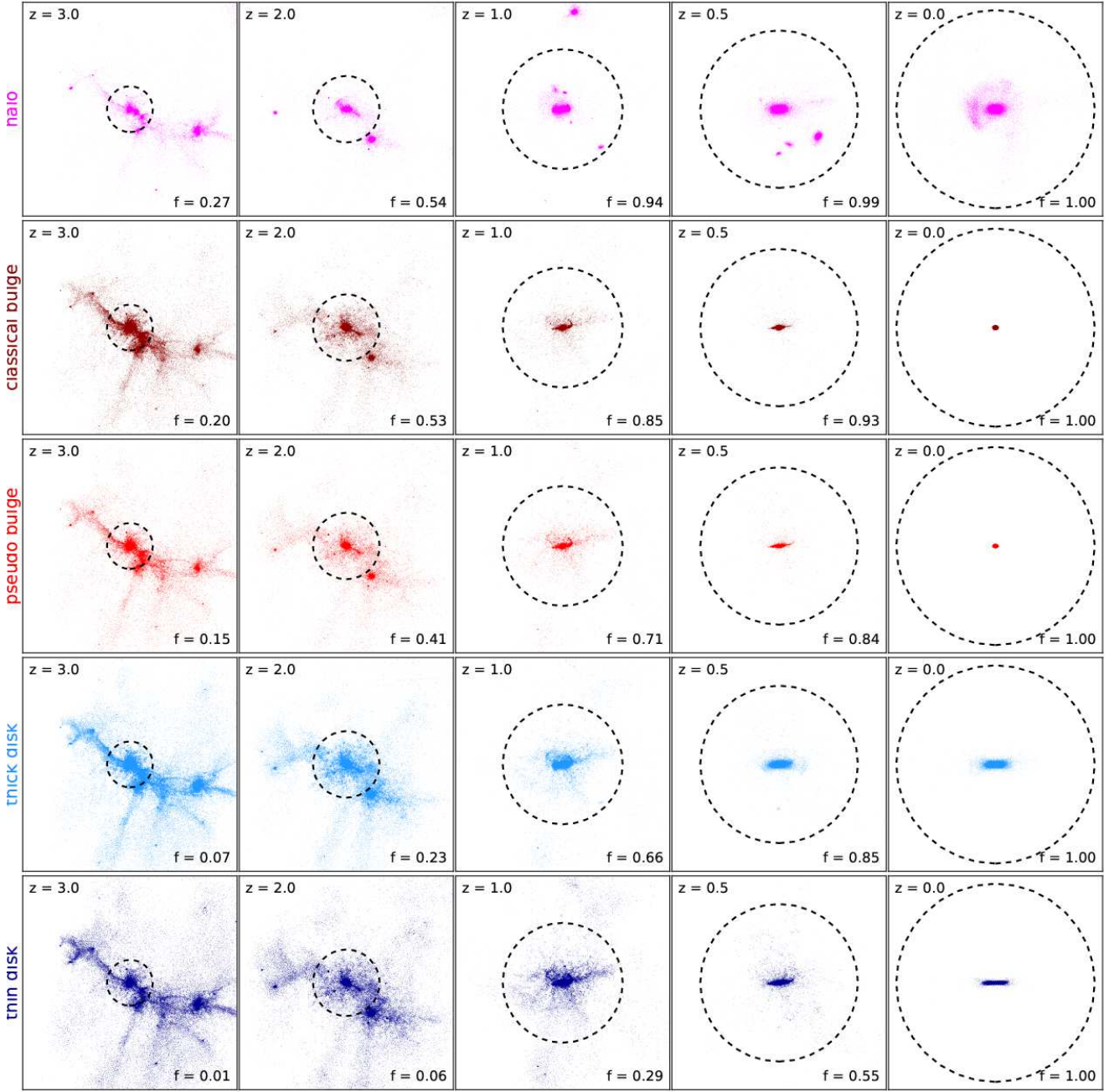}
\caption{The spatial distribution evolution of the progenitors (gas + stars) of the five stellar kinematic components of g8.26e11. 
The dashed circles represent the virial radii at each redshift shown in the top left. The numbers $f$ shown in the 
bottom right corners of each panel give the {corresponding fraction of the total progenitor mass already in stellar form at that particular redshift}.  
All panels are centered on the center of mass of the progenitor dark matter halo at the corresponding redshift. 
The projection is the same across all redshifts, set as the $yz-$plane of $z=0$. The physical scale is $462$~kpc/side.}
\label{g826_xy_ev}
\end{figure*}

\subsection{A visual perspective on the assembly of stellar structures}
\label{ev_vis}

In order to have a visual impression on the different evolutionary paths of the g8.26e11 kinematic components, 
Figure~\ref{g826_xy_ev} shows a redshift sequence (left to right) of the spatial distribution of the baryonic particles comprising each of them (top to bottom). 
The numbers $f$ in the bottom right corner of each panel give the total stellar mass fractions of the particular component shown at each corresponding redshift.
The normalization of $f$ is done with respect to the total baryonic mass of each component separately.
For e.g., at $z=3$ only $f=1\%$ of the thin disc progenitor material has already been converted to stars.
At $z=3$ (left column) the classical bulge material (second row) occupies the densest regions of the large scale filamentary structure feeding the dark matter halo, 
while the thin disc material (bottom row) is the most diffuse one. 
As the evolution proceeds, the thin disc is the last in collapsing, without considering the stellar halo which takes a long time to assemble the stars in the tidally disrupted small satellites. 

Considering these projections together with the evolution of the stellar and baryon fractions in Figure~\ref{figure8}, it appears that both the thin and the thick discs 
form partially from material that comes in as gas part of the large scale baryon filamentary structure which contracts and collapses. 
On the other hand the two bulges and the halo seem to accrete most of their mass through mergers, understood as through coalescence of already formed structures.
However, from Section~\ref{evolution} we know that the star formation occurs mostly at high redshift and completely in-situ for the two bulges (see Section~\ref{evolution}). 
This means that their progenitor material from the infalling smaller galaxy at $z=2$
sinks to the centre of the main galaxy in gaseous form and is subsequently transformed into stars.
The stars of this infalling galaxy end up making part of the two stellar discs at redshift $z=0$, especially the thick one, and of the stellar halo. 

\section{Summary and conclusions}
\label{conclusions}

We use one simulated Milky Way analogue (g8.26e11) from the NIHAO suite of galaxies \citep{Wang:2015} to show how 
Gaussian Mixture Models in the stellar kinematic space of normalized specific angular momentum -- binding energy 
 ($j_{\rm z}/j_{\rm c}$, $j_{\rm p}/j_{\rm c}$,$e/|e|_{\rm max}$) 
can disentangle a large variety of galactic stellar structures.

\subsection{Galactic structure finder}

The analysis pipeline {\tt galactic structure finder} ({\tt gsf}) can be applied to 
any simulated galaxy in equilibrium state to disentangle the fine structure of its stellar distribution 
(thin/thick discs, classical/pseudo bulges, stellar haloes, spheroids, inner discs/bars).
The code calculates the N-body gravitational potential for a halo in isolation
in order to correctly compute the stellar circularities. 
These circularities $j_z/j_c$ together with the 
normalized stellar angular momenta in the equatorial plane of the galaxy $j_p/j_c$ 
and with the normalized stellar binding energies $e/|e|_{\rm max}$ are used as input space for the Gaussian Mixture Models clustering method. 
The only input parameter needed to run {\tt gsf} is the number of Gaussians $nk$. 
The $nk$ parameter depends on the problem one wants to study and on the resolution of the simulation.

We used {\tt gsf} on a sample of 25 high resolution galaxies (~10$^{\rm 6}$ particles per halo) ranging from dwarfs to objects a few times more massive than the Milky Way \citep{Wang:2015}
to study the properties of stellar structures like thin/thick discs, stellar haloes, etc, as well as their formation history. 
In this simulated galaxy sample, the low mass objects have only two dynamically distinct components: a disc and a spheroid. 
Driven by the quest to disentangle stellar haloes, we found the more massive galaxies to host up to five distinct components. 
In the present study we exemplify how {\tt gsf} works on a simulated Milky Way analogue. 

For this study, the so-called optimal number of stellar components has been chosen by visual inspection of the surface mass density and 
line-of-sight velocity maps. Automatizing this part is a work in progress, with the aim of applying {\tt gsf} to large samples of simulated galaxies.

\subsection{The multiple components of a MW analogue}

The example galaxy at $z=0$, g8.26e11 has two distinct discs (thin and thick), two distinct bulges (classical and pseudo), and a stellar halo. 
The stellar mass is approximately distributed as follows: $21\%$ in the thin disc, $33\%$ in the thick one, $25\%$ in the classical bulge, $14.5\%$ in the pseudo one, 
and $6.5\%$ in the stellar halo. Therefore, this galaxy has a dynamical disc--to--total ratio of 0.54. 
By comparison the Milky Way is thought to have a disc-to-total ratio of 0.70 \citep[e.g.][]{Bland-Hawthorn:2016}.

Adopting as vantage point a position similar to the Sun's in the MW, g8.26e11 is remarkably similar to the Galaxy.
The total circular velocity for $R>5$~kpc of this simulated galaxy passes through the observational derived data points for the MW of \citet{Kafle:2012}, 
\citet{Reid:2014} and \citet{LopezCorredoira:2014} (Figure~\ref{figure5}).
The thin and thick discs in the simulations have rotational velocities $V_{\phi}$ at the Sun's position ($R_0=8.2$~kpc) of $218$ and $166$~km~s$^{\rm -1}$, respectively, 
values which are in very good agreement with MW observations \citep[e.g.][]{Haywood:2013}. 
The $V_{\phi}$ of the local stellar halo \citep{Bond:2010} is also well recovered ($V_{\phi}\simeq48$~km~s$^{\rm -1}$).
At $R_0$, the vertical velocity dispersions of the thin and thick discs are $29$ and $73$~km~s$^{\rm -1}$, respectively, 
values close to the upper limits found for the MW \citep{Binney:2012,Robin:2017}.

Seen as an extragalactic object, the kinematic thin disc has a flat rotation profile $v_{\rm los}\simeq174\pm5$~km~s$^{\rm -1}$ 
and a small and approximately constant with radius vertical velocity dispersion $\sigma_z\simeq 27\pm6$~km~s$^{\rm -1}$ (Figure~\ref{figure6}).
Its flattening measured from the eigenvalues of the inertia tensor is $\sim0.85$, where one corresponds to a razor thin disc and zero to a perfect spheroid. 
The thick disc, on the other hand, has a declining rotation curve, lagging $\sim50$~km~s$^{\rm -1}$ behind the thin disc at all radii, 
and a significantly larger velocity dispersion ($\sim80$~km~s$^{\rm -1}$ at an edge-on projected radius of $\sim2.5$~kpc) that declines $\sim$linearly with radius. 
The flattening of the thick disc is $\sim0.65$. 
The galaxy as a whole has a large central vertical velocity dispersion ($\sigma_z\simeq 120$~km~s$^{\rm -1}$) due to the presence of the classical and pseudo bulges. 

The simulated galaxy also nicely exemplifies the differences in the various velocities used to judge simulated and observed galaxies. 
The simulated galaxy in this study has $V_c>V_{\phi}>v_{\rm los}$ at all radii. The stellar component whose $V_{\phi}$ and $v_{\rm los}$ are the closest to the total circular velocity is the thin disc. 
Though the cold gas $V_{\phi}$ of this galaxy traces very well $V_c$ ouside of the bulge region, its $v_{\rm los}$ is significantly lower. 
These findings strongly suggest that circular velocities of external galaxies, constructed from observed velocities corrected for inclination effects, can be significantly underestimated.

Contrary to expectations, both types of discs show stellar surface mass density profiles more centrally concentrated than pure exponentials (Figure~\ref{figure6.1}). 
On the other hand, the surface mass density of the classical and pseudo bulges, and of the stellar halo are exponential. 
The flattening for the classical and pseudo bulges are closer to spherical symmetry, $\sim0.20$ and $\sim0.45$, respectively. 
Basically, the kinematic stellar structure we call \textit{classical bulge} has all the expected properties, except the mass density profile. 
It is compact, almost spherically symmetric, has no net rotation and is made of old stars. 
This finding raises interesting questions regarding the nature of classical bulges, as derived from galaxy photometry, 
which predict high S\'{ersic} indices ($n_{\rm S}>2$) for this type of stellar structures. 

The star formation history and the stellar mass assembly history of this galaxy (Figure~\ref{figure8}) 
are similar to the observational derived ones for a Milky Way mass galaxy \citep{vanDokkum:2013}.
Breaking down the total SFR into the contributions from the five components, we find that the dispersion dominated structures (the two bulges and the stellar halo) 
formed most of their stars at high redshift ($z>1$), the peaks in their SFRs occurring at $z\sim3$. 
The thick disc has a very extended SFR with a quite flat peak ($\sim5$~M$_{\rm\odot}$yr$^{\rm -1}$) between $z\sim2.5$ and $z\sim1.5$, 
while the thin disc forms its stars at a constant rate of $\simeq1.5$~M$_{\rm\odot}$yr$^{\rm -1}$ between $z\simeq2.5$ and $z=0$. 
Globally, this galaxy formed half of its stars by $z_{\rm 1/2}=1.46$. The half stellar mass formation redshifts, $z_{\rm 1/2}$,
for the five structures form a sequence, with $z_{\rm 1/2}=2.15$, 2.13, 1.79, 1.35, and 0.57
for the stellar halo, classical bulge, pseudo bulge, thick disc and thin disc, respectively.

One of the major benefits of our method is that it allows to study the formation histories of these various structures by tracing back in time the Lagrangian mass 
of each one of them separately. Actually, this should be a relatively straightforward analysis in any particle-based simulation code.
In this manner, for example, we can quantify precisely the loss of angular momentum between the epoch of dark matter halo 
turn-around and $z=0$ for each stellar kinematic structure. 
For this particular galaxy, the thin disc material loses the smallest fraction of its maximum angular momentum
($\sim60$ per cent), while the classical bulge loses the most ($\sim95$ per cent). 
Similarly, the $z=0$ dark matter halo only loses $\sim30$ per cent of its maximum angular momentum. 
By plotting the evolution of the various parameters, such as half mass radius, shape, rotational support and/or specific angular momentum, for each stellar kinematic component as well as for 
the dark matter halo, it is possible to identify the important epochs in the formation of the galaxy (Figure~\ref{figure7}). 
In this way we found the turn-around-redshift for this galaxy to be $z_{\rm turn}\sim3$, while the virialisation of its dark matter halo ends by $z_{\rm vir}\sim1.3$.
For all stellar components as well as for the dark matter halo the biggest loss of angular momentum occurs between these two epochs.

At high redshifts all five stellar kinematic components display a filamentary spatial structure, 
which vanishes first for the dispersion dominated structures, and lastly for the rotation dominated ones (Figure~\ref{g826_xy_ev}).
The two bulges of this galaxy formed in-situ all their stars, while the thick disc accretes $8\%$ and forms in-situ the rest. 
The thin disc has also a small fraction of accreted stars of $2\%$. 
The thick stellar disc of this simulated galaxy \textit{forms thick} \citep{Brook:2004}.  
The stellar halo has a different assembly history than the other four components, at least half of its stellar mass being formed in small satellites that subsequently get incorporated
to the progenitor galaxy and are tidally destroyed in this process. A significant fraction ($55\%$) of this galaxy's stellar halo is, however, formed in-situ \citep[e.g.][]{Cooper:2015}. 

\subsection{Ongoing and future {\tt gsf} applications}

Finally, we like to anticipate that in an accompanying paper \citep{Obreja:2018} 
we extend the analysis presented here to a larger set of 25 NIHAO galaxies, 
in a first attempt to constrain the formation patterns of stellar substructures like thin/thick discs, classical/pseudo bulges, stellar haloes, inner stellar discs and stellar spheroids. 

Recent years have seen big advances in the field of high resolution galaxy simulations, resulting in ever more realistic galaxies. 
However, the various groups active in this field use simulation codes which employ different implementations for the sub-grid physics.
Therefore, it is very important to understand what are the detailed differences between these codes in terms of the fine structure of galaxies, 
so that by comparing with observational data the models of galaxy formation can be improved. 
In this perspective, the new generation of zoom-in cosmological simulations with very high resolutions \citep[e.g.][]{Grand:2017,Hopkins:2017,Buck:2018a} 
are an ideal laboratory to study the emergence of the galactic stellar structures.
For these reasons, we think that our analysis pipeline  
would open the path to a better understanding of stellar structures formation if applied in a consistent way 
to the wealth of current and future high-resolution zoom-in simulations.
To foster such studies, we thus make {\tt gsf} publicly available at \url{https://github.com/aobr/gsf}.

\section*{Acknowledgments}

We would like to thank the anonymous referee for a constructive report, which helped improve the quality of this manuscript.
We would also like to thank Glenn van de Ven, Fabrizio Arrigoni Battaia, Rosa Dom\'{\i}nguez Tenreiro and Chris Brook for useful conversations.
All figures in this work have been made with {\tt matplotlib} \citep{Hunter:2007}. 
The {\tt gsf} code also uses the Python libraries {\tt numpy} \citep{Walt:2011} and {\tt scipy} \citep{Jones:2001}. 
{\tt F2PY} \citep{Peterson:2009} has been used to compile the Fortran module for Python.
This research was carried out on the High Performance Computing resources at New York University Abu Dhabi; 
on the \textsc{theo} cluster of the Max-Planck-Institut f\"{u}r Astronomie and on the \textsc{hydra} clusters at the Rechenzentrum in Garching. 
We greatly appreciate the contributions of these computing allocations.
AO and BM have been funded by the Deutsche Forschungsgemeinschaft (DFG, German Research Foundation) -- MO 2979/1-1.
TB acknowledges support from the  Sonderforschungsbereich SFB 881 ``The Milky Way System'' (subproject A1) of the DFG. 

\bibliographystyle{mnras}
\bibliography{paperI}

\begin{thebibliography}{}
\makeatletter
\relax
\def\mn@urlcharsother{\let\do\@makeother \do\$\do\&\do\#\do\^\do\_\do\%\do\~}
\def\mn@doi{\begingroup\mn@urlcharsother \@ifnextchar [ {\mn@doi@}
  {\mn@doi@[]}}
\def\mn@doi@[#1]#2{\def\@tempa{#1}\ifx\@tempa\@empty \href
  {http://dx.doi.org/#2} {doi:#2}\else \href {http://dx.doi.org/#2} {#1}\fi
  \endgroup}
\def\mn@eprint#1#2{\mn@eprint@#1:#2::\@nil}
\def\mn@eprint@arXiv#1{\href {http://arxiv.org/abs/#1} {{\tt arXiv:#1}}}
\def\mn@eprint@dblp#1{\href {http://dblp.uni-trier.de/rec/bibtex/#1.xml}
  {dblp:#1}}
\def\mn@eprint@#1:#2:#3:#4\@nil{\def\@tempa {#1}\def\@tempb {#2}\def\@tempc
  {#3}\ifx \@tempc \@empty \let \@tempc \@tempb \let \@tempb \@tempa \fi \ifx
  \@tempb \@empty \def\@tempb {arXiv}\fi \@ifundefined
  {mn@eprint@\@tempb}{\@tempb:\@tempc}{\expandafter \expandafter \csname
  mn@eprint@\@tempb\endcsname \expandafter{\@tempc}}}

\bibitem[\protect\citeauthoryear{{Abadi}, {Navarro}, {Steinmetz}  \&
  {Eke}}{{Abadi} et~al.}{2003}]{Abadi:2003}
{Abadi} M.~G.,  {Navarro} J.~F.,  {Steinmetz} M.,   {Eke} V.~R.,  2003, \mn@doi
  [\apj] {10.1086/378316}, \href
  {http://adsabs.harvard.edu/abs/2003ApJ...597...21A} {597, 21}

\bibitem[\protect\citeauthoryear{{Aguerri}, {Balcells}  \&
  {Peletier}}{{Aguerri} et~al.}{2001}]{Aguerri:2001}
{Aguerri} J.~A.~L.,  {Balcells} M.,   {Peletier} R.~F.,  2001, \mn@doi [\aap]
  {10.1051/0004-6361:20000441}, \href
  {http://adsabs.harvard.edu/abs/2001A%26A...367..428A} {367, 428}

\bibitem[\protect\citeauthoryear{{Aguerri}, {M{\'e}ndez-Abreu}  \&
  {Corsini}}{{Aguerri} et~al.}{2009}]{Aguerri:2009}
{Aguerri} J.~A.~L.,  {M{\'e}ndez-Abreu} J.,   {Corsini} E.~M.,  2009, \mn@doi
  [\aap] {10.1051/0004-6361:200810931}, \href
  {http://esoads.eso.org/abs/2009A%26A...495..491A} {495, 491}

\bibitem[\protect\citeauthoryear{{Andredakis} \& {Sanders}}{{Andredakis} \&
  {Sanders}}{1994}]{Andredakis:1994}
{Andredakis} Y.~C.,  {Sanders} R.~H.,  1994, \mn@doi [\mnras]
  {10.1093/mnras/267.2.283}, \href
  {http://adsabs.harvard.edu/abs/1994MNRAS.267..283A} {267, 283}

\bibitem[\protect\citeauthoryear{{Andredakis}, {Peletier}  \&
  {Balcells}}{{Andredakis} et~al.}{1995}]{Andredakis:1995}
{Andredakis} Y.~C.,  {Peletier} R.~F.,   {Balcells} M.,  1995, \mn@doi [\mnras]
  {10.1093/mnras/275.3.874}, \href
  {http://adsabs.harvard.edu/abs/1995MNRAS.275..874A} {275, 874}

\bibitem[\protect\citeauthoryear{{Athanassoula}}{{Athanassoula}}{2005}]{Athanassoula:2005}
{Athanassoula} E.,  2005, \mn@doi [\mnras] {10.1111/j.1365-2966.2005.08872.x},
  \href {http://esoads.eso.org/abs/2005MNRAS.358.1477A} {358, 1477}

\bibitem[\protect\citeauthoryear{{Athanassoula}, {Rodionov}, {Peschken}  \&
  {Lambert}}{{Athanassoula} et~al.}{2016}]{Athanassoula:2016}
{Athanassoula} E.,  {Rodionov} S.~A.,  {Peschken} N.,   {Lambert} J.~C.,  2016,
  \mn@doi [\apj] {10.3847/0004-637X/821/2/90}, \href
  {http://adsabs.harvard.edu/abs/2016ApJ...821...90A} {821, 90}

\bibitem[\protect\citeauthoryear{{Aumer}, {White}, {Naab}  \&
  {Scannapieco}}{{Aumer} et~al.}{2013}]{Aumer:2013}
{Aumer} M.,  {White} S.~D.~M.,  {Naab} T.,   {Scannapieco} C.,  2013, \mn@doi
  [\mnras] {10.1093/mnras/stt1230}, \href
  {http://adsabs.harvard.edu/abs/2013MNRAS.434.3142A} {434, 3142}

\bibitem[\protect\citeauthoryear{{Becklin} \& {Neugebauer}}{{Becklin} \&
  {Neugebauer}}{1968}]{Becklin:1968}
{Becklin} E.~E.,  {Neugebauer} G.,  1968, \mn@doi [\apj] {10.1086/149425},
  \href {http://esoads.eso.org/abs/1968ApJ...151..145B} {151, 145}

\bibitem[\protect\citeauthoryear{{Behroozi}, {Wechsler}  \&
  {Conroy}}{{Behroozi} et~al.}{2013}]{Behroozi:2013}
{Behroozi} P.~S.,  {Wechsler} R.~H.,   {Conroy} C.,  2013, \mn@doi [\apj]
  {10.1088/0004-637X/770/1/57}, \href
  {http://adsabs.harvard.edu/abs/2013ApJ...770...57B} {770, 57}

\bibitem[\protect\citeauthoryear{{Bernardi}, {Meert}, {Vikram},
  {Huertas-Company}, {Mei}, {Shankar}  \& {Sheth}}{{Bernardi}
  et~al.}{2014}]{Bernardi:2014}
{Bernardi} M.,  {Meert} A.,  {Vikram} V.,  {Huertas-Company} M.,  {Mei} S.,
  {Shankar} F.,   {Sheth} R.~K.,  2014, \mn@doi [\mnras]
  {10.1093/mnras/stu1106}, \href
  {http://adsabs.harvard.edu/abs/2014MNRAS.443..874B} {443, 874}

\bibitem[\protect\citeauthoryear{{Binney}}{{Binney}}{2012}]{Binney:2012}
{Binney} J.,  2012, \mn@doi [\mnras] {10.1111/j.1365-2966.2012.21692.x}, \href
  {http://adsabs.harvard.edu/abs/2012MNRAS.426.1328B} {426, 1328}

\bibitem[\protect\citeauthoryear{{Binney} \& {Piffl}}{{Binney} \&
  {Piffl}}{2015}]{Binney:2015}
{Binney} J.,  {Piffl} T.,  2015, \mn@doi [\mnras] {10.1093/mnras/stv2225},
  \href {http://adsabs.harvard.edu/abs/2015MNRAS.454.3653B} {454, 3653}

\bibitem[\protect\citeauthoryear{{Bland-Hawthorn} \&
  {Gerhard}}{{Bland-Hawthorn} \& {Gerhard}}{2016}]{Bland-Hawthorn:2016}
{Bland-Hawthorn} J.,  {Gerhard} O.,  2016, \mn@doi [\araa]
  {10.1146/annurev-astro-081915-023441}, \href
  {http://adsabs.harvard.edu/abs/2016ARA%26A..54..529B} {54, 529}

\bibitem[\protect\citeauthoryear{{Bond}, {Ivezi{\'c}}, {Sesar}  \& {et
  al.}}{{Bond} et~al.}{2010}]{Bond:2010}
{Bond} N.~A.,  {Ivezi{\'c}} {\v Z}.,  {Sesar} B.,   {et al.} 2010, \mn@doi
  [\apj] {10.1088/0004-637X/716/1/1}, \href
  {http://adsabs.harvard.edu/abs/2010ApJ...716....1B} {716, 1}

\bibitem[\protect\citeauthoryear{{Bottrell}, {Torrey}, {Simard}  \&
  {Ellison}}{{Bottrell} et~al.}{2017}]{Bottrell:2017}
{Bottrell} C.,  {Torrey} P.,  {Simard} L.,   {Ellison} S.~L.,  2017, \mn@doi
  [\mnras] {10.1093/mnras/stx276}, \href
  {http://adsabs.harvard.edu/abs/2017MNRAS.467.2879B} {467, 2879}

\bibitem[\protect\citeauthoryear{{Brook}, {Kawata}, {Gibson}  \&
  {Freeman}}{{Brook} et~al.}{2004}]{Brook:2004}
{Brook} C.~B.,  {Kawata} D.,  {Gibson} B.~K.,   {Freeman} K.~C.,  2004, \mn@doi
  [\apj] {10.1086/422709}, \href
  {http://adsabs.harvard.edu/abs/2004ApJ...612..894B} {612, 894}

\bibitem[\protect\citeauthoryear{{Brook}, {Stinson}, {Gibson}, {Wadsley}  \&
  {Quinn}}{{Brook} et~al.}{2012}]{Brook:2012}
{Brook} C.~B.,  {Stinson} G.,  {Gibson} B.~K.,  {Wadsley} J.,   {Quinn} T.,
  2012, \mn@doi [\mnras] {10.1111/j.1365-2966.2012.21306.x}, \href
  {http://adsabs.harvard.edu/abs/2012MNRAS.424.1275B} {424, 1275}

\bibitem[\protect\citeauthoryear{{Brooks}}{{Brooks}}{2008}]{Brooks:2008}
{Brooks} A.,  2008, PhD thesis, University of Washington

\bibitem[\protect\citeauthoryear{{Buck}, {Macci{\`o}}, {Obreja}, {Dutton},
  {Dom{\'{\i}}nguez-Tenreiro}  \& {Granato}}{{Buck} et~al.}{2017}]{Buck:2017}
{Buck} T.,  {Macci{\`o}} A.~V.,  {Obreja} A.,  {Dutton} A.~A.,
  {Dom{\'{\i}}nguez-Tenreiro} R.,   {Granato} G.~L.,  2017, \mn@doi [\mnras]
  {10.1093/mnras/stx685}, \href
  {http://adsabs.harvard.edu/abs/2017MNRAS.468.3628B} {468, 3628}

\bibitem[\protect\citeauthoryear{{Buck}, {Macci{\`o}}, {Dutton}, {Obreja}  \&
  {Frings}}{{Buck} et~al.}{2018}]{Buck:2018a}
{Buck} T.,  {Macci{\`o}} A.~V.,  {Dutton} A.~A.,  {Obreja} A.,   {Frings} J.,
  2018, preprint, \href {http://adsabs.harvard.edu/abs/2018arXiv180404667B} {}
  (\mn@eprint {arXiv} {1804.04667})

\bibitem[\protect\citeauthoryear{{Carollo}, {Beers}, {Lee}  \& {et
  al.}}{{Carollo} et~al.}{2007}]{Carollo:2007}
{Carollo} D.,  {Beers} T.~C.,  {Lee} Y.~S.,   {et al.} 2007, \mn@doi [\nat]
  {10.1038/nature06460}, \href
  {http://adsabs.harvard.edu/abs/2007Natur.450.1020C} {450, 1020}

\bibitem[\protect\citeauthoryear{{Chabrier}}{{Chabrier}}{2003}]{Chabrier:2003}
{Chabrier} G.,  2003, \mn@doi [\pasp] {10.1086/376392}, \href
  {http://adsabs.harvard.edu/abs/2003PASP..115..763C} {115, 763}

\bibitem[\protect\citeauthoryear{{Chandrasekhar}}{{Chandrasekhar}}{1943}]{Chandrasekhar:1943}
{Chandrasekhar} S.,  1943, \mn@doi [\apj] {10.1086/144517}, \href
  {http://adsabs.harvard.edu/abs/1943ApJ....97..255C} {97, 255}

\bibitem[\protect\citeauthoryear{{Christensen}, {Brooks}, {Fisher},
  {Governato}, {McCleary}, {Quinn}, {Shen}  \& {Wadsley}}{{Christensen}
  et~al.}{2014}]{Christensen:2014}
{Christensen} C.~R.,  {Brooks} A.~M.,  {Fisher} D.~B.,  {Governato} F.,
  {McCleary} J.,  {Quinn} T.~R.,  {Shen} S.,   {Wadsley} J.,  2014, \mn@doi
  [\mnras] {10.1093/mnrasl/slu020}, \href
  {http://adsabs.harvard.edu/abs/2014MNRAS.440L..51C} {440, L51}

\bibitem[\protect\citeauthoryear{{Comer{\'o}n}, {Elmegreen}, {Knapen}  \& {et
  al.}}{{Comer{\'o}n} et~al.}{2011}]{Comeron:2011}
{Comer{\'o}n} S.,  {Elmegreen} B.~G.,  {Knapen} J.~H.,   {et al.} 2011, \mn@doi
  [\apj] {10.1088/0004-637X/741/1/28}, \href
  {http://adsabs.harvard.edu/abs/2011ApJ...741...28C} {741, 28}

\bibitem[\protect\citeauthoryear{{Comer{\'o}n}, {Elmegreen}, {Salo},
  {Laurikainen}, {Holwerda}  \& {Knapen}}{{Comer{\'o}n}
  et~al.}{2014}]{Comeron:2014}
{Comer{\'o}n} S.,  {Elmegreen} B.~G.,  {Salo} H.,  {Laurikainen} E.,
  {Holwerda} B.~W.,   {Knapen} J.~H.,  2014, \mn@doi [\aap]
  {10.1051/0004-6361/201424412}, \href
  {http://adsabs.harvard.edu/abs/2014A%26A...571A..58C} {571, A58}

\bibitem[\protect\citeauthoryear{{Cooper}, {Parry}, {Lowing}, {Cole}  \&
  {Frenk}}{{Cooper} et~al.}{2015}]{Cooper:2015}
{Cooper} A.~P.,  {Parry} O.~H.,  {Lowing} B.,  {Cole} S.,   {Frenk} C.,  2015,
  \mn@doi [\mnras] {10.1093/mnras/stv2057}, \href
  {http://adsabs.harvard.edu/abs/2015MNRAS.454.3185C} {454, 3185}

\bibitem[\protect\citeauthoryear{{Dalcanton} \& {Bernstein}}{{Dalcanton} \&
  {Bernstein}}{2002}]{Dalcanton:2002}
{Dalcanton} J.~J.,  {Bernstein} R.~A.,  2002, \mn@doi [\aj] {10.1086/342286},
  \href {http://adsabs.harvard.edu/abs/2002AJ....124.1328D} {124, 1328}

\bibitem[\protect\citeauthoryear{{Davies} \& {Illingworth}}{{Davies} \&
  {Illingworth}}{1983}]{Davies:1983}
{Davies} R.~L.,  {Illingworth} G.,  1983, \mn@doi [\apj] {10.1086/160799},
  \href {http://adsabs.harvard.edu/abs/1983ApJ...266..516D} {266, 516}

\bibitem[\protect\citeauthoryear{{Dehnen} \& {Aly}}{{Dehnen} \&
  {Aly}}{2012}]{Dehnen:2012}
{Dehnen} W.,  {Aly} H.,  2012, \mn@doi [\mnras]
  {10.1111/j.1365-2966.2012.21439.x}, \href
  {http://adsabs.harvard.edu/abs/2012MNRAS.425.1068D} {425, 1068}

\bibitem[\protect\citeauthoryear{{Dom{\'e}nech-Moral},
  {Mart{\'{\i}}nez-Serrano}, {Dom{\'{\i}}nguez-Tenreiro}  \&
  {Serna}}{{Dom{\'e}nech-Moral} et~al.}{2012}]{Domenech:2012}
{Dom{\'e}nech-Moral} M.,  {Mart{\'{\i}}nez-Serrano} F.~J.,
  {Dom{\'{\i}}nguez-Tenreiro} R.,   {Serna} A.,  2012, \mn@doi [\mnras]
  {10.1111/j.1365-2966.2012.20534.x}, \href
  {http://adsabs.harvard.edu/abs/2012MNRAS.tmp.2503D} {p.~2503}

\bibitem[\protect\citeauthoryear{{Dom{\'{\i}}nguez-Tenreiro}, {Obreja},
  {Brook}, {Mart{\'{\i}}nez-Serrano}, {Stinson}  \&
  {Serna}}{{Dom{\'{\i}}nguez-Tenreiro} et~al.}{2015}]{Dominguez:2015}
{Dom{\'{\i}}nguez-Tenreiro} R.,  {Obreja} A.,  {Brook} C.~B.,
  {Mart{\'{\i}}nez-Serrano} F.~J.,  {Stinson} G.,   {Serna} A.,  2015, \mn@doi
  [\apjl] {10.1088/2041-8205/800/2/L30}, \href
  {http://adsabs.harvard.edu/abs/2015ApJ...800L..30D} {800, L30}

\bibitem[\protect\citeauthoryear{{Doroshkevich}}{{Doroshkevich}}{1970}]{Doroshkevich:1970}
{Doroshkevich} A.~G.,  1970, Astrofizika, \href
  {http://adsabs.harvard.edu/abs/1970Afz.....6..581D} {6, 581}

\bibitem[\protect\citeauthoryear{{Dutton}, {Obreja}, {Wang}  \& {et
  al.}}{{Dutton} et~al.}{2017}]{Dutton:2017}
{Dutton} A.~A.,  {Obreja} A.,  {Wang} L.,   {et al.} 2017, \mn@doi [\mnras]
  {10.1093/mnras/stx458}, \href
  {http://adsabs.harvard.edu/abs/2017MNRAS.467.4937D} {467, 4937}

\bibitem[\protect\citeauthoryear{{Eggen}, {Lynden-Bell}  \& {Sandage}}{{Eggen}
  et~al.}{1962}]{Eggen:1962}
{Eggen} O.~J.,  {Lynden-Bell} D.,   {Sandage} A.~R.,  1962, \mn@doi [\apj]
  {10.1086/147433}, \href {http://adsabs.harvard.edu/abs/1962ApJ...136..748E}
  {136, 748}

\bibitem[\protect\citeauthoryear{{Elmegreen}, {Elmegreen}, {Tompkins}  \&
  {Jenks}}{{Elmegreen} et~al.}{2017}]{Elmegreen:2017}
{Elmegreen} B.~G.,  {Elmegreen} D.~M.,  {Tompkins} B.,   {Jenks} L.~G.,  2017,
  \mn@doi [\apj] {10.3847/1538-4357/aa88d4}, \href
  {http://adsabs.harvard.edu/abs/2017ApJ...847...14E} {847, 14}

\bibitem[\protect\citeauthoryear{Eric~Jones et~al.}{Eric~Jones
  et~al.}{2001}]{Jones:2001}
Eric~Jones Travis~Oliphant P.~P.,  et~al., 2001, {SciPy: Open source scientific
  tools for Python}, \url {http://www.scipy.org/}

\bibitem[\protect\citeauthoryear{{Ferrarese} \& {Merritt}}{{Ferrarese} \&
  {Merritt}}{2000}]{Ferrarese:2000}
{Ferrarese} L.,  {Merritt} D.,  2000, \mn@doi [\apjl] {10.1086/312838}, \href
  {http://esoads.eso.org/abs/2000ApJ...539L...9F} {539, L9}

\bibitem[\protect\citeauthoryear{{Fisher} \& {Drory}}{{Fisher} \&
  {Drory}}{2008}]{Fisher:2008}
{Fisher} D.~B.,  {Drory} N.,  2008, \mn@doi [\aj]
  {10.1088/0004-6256/136/2/773}, \href
  {http://adsabs.harvard.edu/abs/2008AJ....136..773F} {136, 773}

\bibitem[\protect\citeauthoryear{{Gadotti}}{{Gadotti}}{2009}]{Gadotti:2009}
{Gadotti} D.~A.,  2009, \mn@doi [\mnras] {10.1111/j.1365-2966.2008.14257.x},
  \href {http://adsabs.harvard.edu/abs/2009MNRAS.393.1531G} {393, 1531}

\bibitem[\protect\citeauthoryear{{Gebhardt}, {Bender}, {Bower}  \& {et.
  al}}{{Gebhardt} et~al.}{2000}]{Gebhardt:2000}
{Gebhardt} K.,  {Bender} R.,  {Bower} G.,   {et. al} 2000, \mn@doi [\apjl]
  {10.1086/312840}, \href {http://esoads.eso.org/abs/2000ApJ...539L..13G} {539,
  L13}

\bibitem[\protect\citeauthoryear{{Gilmore} \& {Reid}}{{Gilmore} \&
  {Reid}}{1983}]{Gilmore:1983}
{Gilmore} G.,  {Reid} N.,  1983, \mn@doi [\mnras] {10.1093/mnras/202.4.1025},
  \href {http://esoads.eso.org/abs/1983MNRAS.202.1025G} {202, 1025}

\bibitem[\protect\citeauthoryear{{Gonz{\'a}lez-Garc{\'{\i}}a} \& {van
  Albada}}{{Gonz{\'a}lez-Garc{\'{\i}}a} \& {van
  Albada}}{2005}]{GonzalezGarcia:2005}
{Gonz{\'a}lez-Garc{\'{\i}}a} A.~C.,  {van Albada} T.~S.,  2005, \mn@doi
  [\mnras] {10.1111/j.1365-2966.2005.09242.x}, \href
  {http://adsabs.harvard.edu/abs/2005MNRAS.361.1030G} {361, 1030}

\bibitem[\protect\citeauthoryear{{Governato}, {Mayer}, {Wadsley}  \& {et
  al.}}{{Governato} et~al.}{2004}]{Governato:2004}
{Governato} F.,  {Mayer} L.,  {Wadsley} J.,   {et al.} 2004, \mn@doi [\apj]
  {10.1086/383516}, \href {http://adsabs.harvard.edu/abs/2004ApJ...607..688G}
  {607, 688}

\bibitem[\protect\citeauthoryear{{Graham} \& {Guzm{\'a}n}}{{Graham} \&
  {Guzm{\'a}n}}{2003}]{Graham:2003}
{Graham} A.~W.,  {Guzm{\'a}n} R.,  2003, \mn@doi [\aj] {10.1086/374992}, \href
  {http://adsabs.harvard.edu/abs/2003AJ....125.2936G} {125, 2936}

\bibitem[\protect\citeauthoryear{{Grand} et~al.,}{{Grand}
  et~al.}{2017}]{Grand:2017}
{Grand} R.~J.~J.,  et~al., 2017, \mn@doi [\mnras] {10.1093/mnras/stx071}, \href
  {http://adsabs.harvard.edu/abs/2017MNRAS.467..179G} {467, 179}

\bibitem[\protect\citeauthoryear{{Guidi}, {Scannapieco}, {Walcher}  \&
  {Gallazzi}}{{Guidi} et~al.}{2016}]{Guidi:2016}
{Guidi} G.,  {Scannapieco} C.,  {Walcher} J.,   {Gallazzi} A.,  2016, \mn@doi
  [\mnras] {10.1093/mnras/stw1790}, \href
  {http://adsabs.harvard.edu/abs/2016MNRAS.462.2046G} {462, 2046}

\bibitem[\protect\citeauthoryear{{Haardt} \& {Madau}}{{Haardt} \&
  {Madau}}{2012}]{Haardt:2012}
{Haardt} F.,  {Madau} P.,  2012, \mn@doi [\apj] {10.1088/0004-637X/746/2/125},
  \href {http://adsabs.harvard.edu/abs/2012ApJ...746..125H} {746, 125}

\bibitem[\protect\citeauthoryear{{Hammersley}, {Garz{\'o}n}, {Mahoney},
  {L{\'o}pez-Corredoira}  \& {Torres}}{{Hammersley}
  et~al.}{2000}]{Hammersley:2000}
{Hammersley} P.~L.,  {Garz{\'o}n} F.,  {Mahoney} T.~J.,  {L{\'o}pez-Corredoira}
  M.,   {Torres} M.~A.~P.,  2000, \mn@doi [\mnras]
  {10.1046/j.1365-8711.2000.03858.x}, \href
  {http://esoads.eso.org/abs/2000MNRAS.317L..45H} {317, L45}

\bibitem[\protect\citeauthoryear{{H{\"a}ring} \& {Rix}}{{H{\"a}ring} \&
  {Rix}}{2004}]{Haring:2004}
{H{\"a}ring} N.,  {Rix} H.-W.,  2004, \mn@doi [\apjl] {10.1086/383567}, \href
  {http://adsabs.harvard.edu/abs/2004ApJ...604L..89H} {604, L89}

\bibitem[\protect\citeauthoryear{{Haywood}, {Di Matteo}, {Lehnert}, {Katz}  \&
  {G{\'o}mez}}{{Haywood} et~al.}{2013}]{Haywood:2013}
{Haywood} M.,  {Di Matteo} P.,  {Lehnert} M.~D.,  {Katz} D.,   {G{\'o}mez} A.,
  2013, \mn@doi [\aap] {10.1051/0004-6361/201321397}, \href
  {http://adsabs.harvard.edu/abs/2013A%26A...560A.109H} {560, A109}

\bibitem[\protect\citeauthoryear{{Hopkins}, {Kere{\v s}}, {O{\~n}orbe},
  {Faucher-Gigu{\`e}re}, {Quataert}, {Murray}  \& {Bullock}}{{Hopkins}
  et~al.}{2014}]{Hopkins:2014}
{Hopkins} P.~F.,  {Kere{\v s}} D.,  {O{\~n}orbe} J.,  {Faucher-Gigu{\`e}re}
  C.-A.,  {Quataert} E.,  {Murray} N.,   {Bullock} J.~S.,  2014, \mn@doi
  [\mnras] {10.1093/mnras/stu1738}, \href
  {http://adsabs.harvard.edu/abs/2014MNRAS.445..581H} {445, 581}

\bibitem[\protect\citeauthoryear{{Hopkins}, {Wetzel}, {Keres}  \& {et
  al.}}{{Hopkins} et~al.}{2017}]{Hopkins:2017}
{Hopkins} P.~F.,  {Wetzel} A.,  {Keres} D.,   {et al.} 2017, preprint, \href
  {http://adsabs.harvard.edu/abs/2017arXiv170206148H} {} (\mn@eprint {arXiv}
  {1702.06148})

\bibitem[\protect\citeauthoryear{{Hoyle}}{{Hoyle}}{1951}]{Hoyle:1951}
{Hoyle} F.,  1951, in Problems of Cosmical Aerodynamics. p.~195

\bibitem[\protect\citeauthoryear{{Hunter}}{{Hunter}}{2007}]{Hunter:2007}
{Hunter} J.~D.,  2007, \mn@doi [Computing in Science and Engineering]
  {10.1109/MCSE.2007.55}, \href
  {http://adsabs.harvard.edu/abs/2007CSE.....9...90H} {9, 90}

\bibitem[\protect\citeauthoryear{{Jahnke} \& {Macci{\`o}}}{{Jahnke} \&
  {Macci{\`o}}}{2011}]{Jahnke:2011}
{Jahnke} K.,  {Macci{\`o}} A.~V.,  2011, \mn@doi [\apj]
  {10.1088/0004-637X/734/2/92}, \href
  {http://adsabs.harvard.edu/abs/2011ApJ...734...92J} {734, 92}

\bibitem[\protect\citeauthoryear{{Kafle}, {Sharma}, {Lewis}  \&
  {Bland-Hawthorn}}{{Kafle} et~al.}{2012}]{Kafle:2012}
{Kafle} P.~R.,  {Sharma} S.,  {Lewis} G.~F.,   {Bland-Hawthorn} J.,  2012,
  \mn@doi [\apj] {10.1088/0004-637X/761/2/98}, \href
  {http://adsabs.harvard.edu/abs/2012ApJ...761...98K} {761, 98}

\bibitem[\protect\citeauthoryear{{Kannan}, {Macci{\`o}}, {Fontanot}, {Moster},
  {Karman}  \& {Somerville}}{{Kannan} et~al.}{2015}]{Kannan:2015}
{Kannan} R.,  {Macci{\`o}} A.~V.,  {Fontanot} F.,  {Moster} B.~P.,  {Karman}
  W.,   {Somerville} R.~S.,  2015, \mn@doi [\mnras] {10.1093/mnras/stv1633},
  \href {http://adsabs.harvard.edu/abs/2015MNRAS.452.4347K} {452, 4347}

\bibitem[\protect\citeauthoryear{{Knollmann} \& {Knebe}}{{Knollmann} \&
  {Knebe}}{2009}]{Knollmann:2009}
{Knollmann} S.~R.,  {Knebe} A.,  2009, \mn@doi [\apjs]
  {10.1088/0067-0049/182/2/608}, \href
  {http://adsabs.harvard.edu/abs/2009ApJS..182..608K} {182, 608}

\bibitem[\protect\citeauthoryear{{Koda}, {Yagi}, {Yamanoi}  \&
  {Komiyama}}{{Koda} et~al.}{2015}]{Koda:2015}
{Koda} J.,  {Yagi} M.,  {Yamanoi} H.,   {Komiyama} Y.,  2015, \mn@doi [\apjl]
  {10.1088/2041-8205/807/1/L2}, \href
  {http://adsabs.harvard.edu/abs/2015ApJ...807L...2K} {807, L2}

\bibitem[\protect\citeauthoryear{{Kormendy} \& {Barentine}}{{Kormendy} \&
  {Barentine}}{2010}]{Kormendy:2010}
{Kormendy} J.,  {Barentine} J.~C.,  2010, \mn@doi [\apjl]
  {10.1088/2041-8205/715/2/L176}, \href
  {http://esoads.eso.org/abs/2010ApJ...715L.176K} {715, L176}

\bibitem[\protect\citeauthoryear{{Kormendy} \& {Kennicutt}}{{Kormendy} \&
  {Kennicutt}}{2004}]{Kormendy:2004}
{Kormendy} J.,  {Kennicutt} Jr. R.~C.,  2004, \mn@doi [\araa]
  {10.1146/annurev.astro.42.053102.134024}, \href
  {http://adsabs.harvard.edu/abs/2004ARA%26A..42..603K} {42, 603}

\bibitem[\protect\citeauthoryear{{Leeuwin} \& {Combes}}{{Leeuwin} \&
  {Combes}}{1997}]{Leeuwin:1997}
{Leeuwin} F.,  {Combes} F.,  1997, \mn@doi [\mnras] {10.1093/mnras/284.1.45},
  \href {http://adsabs.harvard.edu/abs/1997MNRAS.284...45L} {284, 45}

\bibitem[\protect\citeauthoryear{{L{\'o}pez-Corredoira}}{{L{\'o}pez-Corredoira}}{2014}]{LopezCorredoira:2014}
{L{\'o}pez-Corredoira} M.,  2014, \mn@doi [\aap] {10.1051/0004-6361/201423505},
  \href {http://adsabs.harvard.edu/abs/2014A%26A...563A.128L} {563, A128}

\bibitem[\protect\citeauthoryear{{Lynden-Bell}}{{Lynden-Bell}}{1967}]{LyndenBell:1967}
{Lynden-Bell} D.,  1967, \mn@doi [\mnras] {10.1093/mnras/136.1.101}, \href
  {http://adsabs.harvard.edu/abs/1967MNRAS.136..101L} {136, 101}

\bibitem[\protect\citeauthoryear{{Lynden-Bell} \& {Kalnajs}}{{Lynden-Bell} \&
  {Kalnajs}}{1972}]{Lynden-Bell:1972}
{Lynden-Bell} D.,  {Kalnajs} A.~J.,  1972, \mn@doi [\mnras]
  {10.1093/mnras/157.1.1}, \href
  {http://esoads.eso.org/abs/1972MNRAS.157....1L} {157, 1}

\bibitem[\protect\citeauthoryear{{Marinacci}, {Pakmor}  \&
  {Springel}}{{Marinacci} et~al.}{2014}]{Marinacci:2014}
{Marinacci} F.,  {Pakmor} R.,   {Springel} V.,  2014, \mn@doi [\mnras]
  {10.1093/mnras/stt2003}, \href
  {http://adsabs.harvard.edu/abs/2014MNRAS.437.1750M} {437, 1750}

\bibitem[\protect\citeauthoryear{{Martig}, {Bournaud}, {Croton}, {Dekel}  \&
  {Teyssier}}{{Martig} et~al.}{2012}]{Martig:2012}
{Martig} M.,  {Bournaud} F.,  {Croton} D.~J.,  {Dekel} A.,   {Teyssier} R.,
  2012, \mn@doi [\apj] {10.1088/0004-637X/756/1/26}, \href
  {http://adsabs.harvard.edu/abs/2012ApJ...756...26M} {756, 26}

\bibitem[\protect\citeauthoryear{{Martinsson}, {Verheijen}, {Westfall},
  {Bershady}, {Schechtman-Rook}, {Andersen}  \& {Swaters}}{{Martinsson}
  et~al.}{2013}]{Martinsson:2013}
{Martinsson} T.~P.~K.,  {Verheijen} M.~A.~W.,  {Westfall} K.~B.,  {Bershady}
  M.~A.,  {Schechtman-Rook} A.,  {Andersen} D.~R.,   {Swaters} R.~A.,  2013,
  \mn@doi [\aap] {10.1051/0004-6361/201220515}, \href
  {http://adsabs.harvard.edu/abs/2013A%26A...557A.130M} {557, A130}

\bibitem[\protect\citeauthoryear{{M{\'e}ndez-Abreu}, {Debattista}, {Corsini}
  \& {Aguerri}}{{M{\'e}ndez-Abreu} et~al.}{2014}]{Mendez-Abreu:2014}
{M{\'e}ndez-Abreu} J.,  {Debattista} V.~P.,  {Corsini} E.~M.,   {Aguerri}
  J.~A.~L.,  2014, \mn@doi [\aap] {10.1051/0004-6361/201423955}, \href
  {http://esoads.eso.org/abs/2014A%26A...572A..25M} {572, A25}

\bibitem[\protect\citeauthoryear{{Mosenkov}, {Sotnikova}  \&
  {Reshetnikov}}{{Mosenkov} et~al.}{2014}]{Mosenkov:2014}
{Mosenkov} A.~V.,  {Sotnikova} N.~Y.,   {Reshetnikov} V.~P.,  2014, \mn@doi
  [\mnras] {10.1093/mnras/stu602}, \href
  {http://adsabs.harvard.edu/abs/2014MNRAS.441.1066M} {441, 1066}

\bibitem[\protect\citeauthoryear{{Moster}, {Naab}  \& {White}}{{Moster}
  et~al.}{2013}]{Moster:2013}
{Moster} B.~P.,  {Naab} T.,   {White} S.~D.~M.,  2013, \mn@doi [\mnras]
  {10.1093/mnras/sts261}, \href
  {http://adsabs.harvard.edu/abs/2013MNRAS.428.3121M} {428, 3121}

\bibitem[\protect\citeauthoryear{{Navarro} \& {Steinmetz}}{{Navarro} \&
  {Steinmetz}}{2000}]{Navarro:2000}
{Navarro} J.~F.,  {Steinmetz} M.,  2000, \mn@doi [\apj] {10.1086/309175}, \href
  {http://adsabs.harvard.edu/abs/2000ApJ...538..477N} {538, 477}

\bibitem[\protect\citeauthoryear{{Nowak}, {Thomas}, {Erwin}, {Saglia}, {Bender}
   \& {Davies}}{{Nowak} et~al.}{2010}]{Nowak:2010}
{Nowak} N.,  {Thomas} J.,  {Erwin} P.,  {Saglia} R.~P.,  {Bender} R.,
  {Davies} R.~I.,  2010, \mn@doi [\mnras] {10.1111/j.1365-2966.2009.16167.x},
  \href {http://esoads.eso.org/abs/2010MNRAS.403..646N} {403, 646}

\bibitem[\protect\citeauthoryear{{Obreja}, {Brook}, {Stinson},
  {Dom{\'{\i}}nguez-Tenreiro}, {Gibson}, {Silva}  \& {Granato}}{{Obreja}
  et~al.}{2014}]{Obreja:2014}
{Obreja} A.,  {Brook} C.~B.,  {Stinson} G.,  {Dom{\'{\i}}nguez-Tenreiro} R.,
  {Gibson} B.~K.,  {Silva} L.,   {Granato} G.~L.,  2014, \mn@doi [\mnras]
  {10.1093/mnras/stu891}, \href
  {http://adsabs.harvard.edu/abs/2014MNRAS.442.1794O} {442, 1794}

\bibitem[\protect\citeauthoryear{{Obreja}, {Stinson}, {Dutton}, {Macci{\`o}},
  {Wang}  \& {Kang}}{{Obreja} et~al.}{2016}]{Obreja:2016}
{Obreja} A.,  {Stinson} G.~S.,  {Dutton} A.~A.,  {Macci{\`o}} A.~V.,  {Wang}
  L.,   {Kang} X.,  2016, \mn@doi [\mnras] {10.1093/mnras/stw690}, \href
  {http://adsabs.harvard.edu/abs/2016MNRAS.459..467O} {459, 467}

\bibitem[\protect\citeauthoryear{{Obreja}, {Dutton}, {Macci{\`o}}  \& {et
  al.}}{{Obreja} et~al.}{2018}]{Obreja:2018}
{Obreja} A.,  {Dutton} A.~A.,  {Macci{\`o}} A.~V.,   {et al.} 2018, preprint,
  \href {http://adsabs.harvard.edu/abs/2018arXiv180406635O} {} (\mn@eprint
  {arXiv} {1804.06635})

\bibitem[\protect\citeauthoryear{{Okamoto}, {Eke}, {Frenk}  \&
  {Jenkins}}{{Okamoto} et~al.}{2005}]{Okamoto:2005}
{Okamoto} T.,  {Eke} V.~R.,  {Frenk} C.~S.,   {Jenkins} A.,  2005, \mn@doi
  [\mnras] {10.1111/j.1365-2966.2005.09525.x}, \href
  {http://adsabs.harvard.edu/abs/2005MNRAS.363.1299O} {363, 1299}

\bibitem[\protect\citeauthoryear{{Okuda}, {Maihara}, {Oda}  \&
  {Sugiyama}}{{Okuda} et~al.}{1977}]{Okuda:1977}
{Okuda} H.,  {Maihara} T.,  {Oda} N.,   {Sugiyama} T.,  1977, \mn@doi [\nat]
  {10.1038/265515a0}, \href {http://adsabs.harvard.edu/abs/1977Natur.265..515O}
  {265, 515}

\bibitem[\protect\citeauthoryear{Pedregosa, Varoquaux, Gramfort  \& {et
  al.}}{Pedregosa et~al.}{2011}]{Pedregosa:2011}
Pedregosa F.,  Varoquaux G.,  Gramfort A.,   {et al.} 2011, Journal of Machine
  Learning Research, 12, 2825

\bibitem[\protect\citeauthoryear{{Peebles}}{{Peebles}}{1969}]{Peebles:1969}
{Peebles} P.~J.~E.,  1969, \mn@doi [\apj] {10.1086/149876}, \href
  {http://adsabs.harvard.edu/abs/1969ApJ...155..393P} {155, 393}

\bibitem[\protect\citeauthoryear{{Peschken}, {Athanassoula}  \&
  {Rodionov}}{{Peschken} et~al.}{2017}]{Peschken:2017}
{Peschken} N.,  {Athanassoula} E.,   {Rodionov} S.~A.,  2017, \mn@doi [\mnras]
  {10.1093/mnras/stx481}, \href
  {http://adsabs.harvard.edu/abs/2017MNRAS.468..994P} {468, 994}

\bibitem[\protect\citeauthoryear{Peterson}{Peterson}{2009}]{Peterson:2009}
Peterson P.,  2009, International Journal of Computational Science and
  Engineering, 4, 296

\bibitem[\protect\citeauthoryear{{Piffl}, {Binney}, {McMillan}  \& {et
  al.}}{{Piffl} et~al.}{2014}]{Piffl:2014}
{Piffl} T.,  {Binney} J.,  {McMillan} P.~J.,   {et al.} 2014, \mn@doi [\mnras]
  {10.1093/mnras/stu1948}, \href
  {http://adsabs.harvard.edu/abs/2014MNRAS.445.3133P} {445, 3133}

\bibitem[\protect\citeauthoryear{{Pontzen}, {Ro{\v s}kar}, {Stinson}  \&
  {Woods}}{{Pontzen} et~al.}{2013}]{Pontzen:2013}
{Pontzen} A.,  {Ro{\v s}kar} R.,  {Stinson} G.,   {Woods} R.,  2013, {pynbody:
  N-Body/SPH analysis for python}, Astrophysics Source Code Library (\mn@eprint
  {ascl} {1305.002})

\bibitem[\protect\citeauthoryear{{Price}}{{Price}}{2008}]{Price:2008}
{Price} D.~J.,  2008, \mn@doi [Journal of Computational Physics]
  {10.1016/j.jcp.2008.08.011}, \href
  {http://adsabs.harvard.edu/abs/2008JCoPh.22710040P} {227, 10040}

\bibitem[\protect\citeauthoryear{{Reid}, {Menten}, {Brunthaler}  \& {et
  al.}}{{Reid} et~al.}{2014}]{Reid:2014}
{Reid} M.~J.,  {Menten} K.~M.,  {Brunthaler} A.,   {et al.} 2014, \mn@doi
  [\apj] {10.1088/0004-637X/783/2/130}, \href
  {http://adsabs.harvard.edu/abs/2014ApJ...783..130R} {783, 130}

\bibitem[\protect\citeauthoryear{{Ritchie} \& {Thomas}}{{Ritchie} \&
  {Thomas}}{2001}]{Ritchie:2001}
{Ritchie} B.~W.,  {Thomas} P.~A.,  2001, \mn@doi [\mnras]
  {10.1046/j.1365-8711.2001.04268.x}, \href
  {http://adsabs.harvard.edu/abs/2001MNRAS.323..743R} {323, 743}

\bibitem[\protect\citeauthoryear{{Robin}, {Bienaym{\'e}},
  {Fern{\'a}ndez-Trincado}  \& {Reyl{\'e}}}{{Robin} et~al.}{2017}]{Robin:2017}
{Robin} A.~C.,  {Bienaym{\'e}} O.,  {Fern{\'a}ndez-Trincado} J.~G.,
  {Reyl{\'e}} C.,  2017, \mn@doi [\aap] {10.1051/0004-6361/201630217}, \href
  {http://adsabs.harvard.edu/abs/2017A%26A...605A...1R} {605, A1}

\bibitem[\protect\citeauthoryear{{Rubin}, {Ford}  \& {Thonnard}}{{Rubin}
  et~al.}{1980}]{Rubin:1980}
{Rubin} V.~C.,  {Ford} Jr. W.~K.,   {Thonnard} N.,  1980, \mn@doi [\apj]
  {10.1086/158003}, \href {http://adsabs.harvard.edu/abs/1980ApJ...238..471R}
  {238, 471}

\bibitem[\protect\citeauthoryear{{Saitoh} \& {Makino}}{{Saitoh} \&
  {Makino}}{2009}]{Saitoh:2009}
{Saitoh} T.~R.,  {Makino} J.,  2009, \mn@doi [\apjl]
  {10.1088/0004-637X/697/2/L99}, \href
  {http://adsabs.harvard.edu/abs/2009ApJ...697L..99S} {697, L99}

\bibitem[\protect\citeauthoryear{{Sandage}}{{Sandage}}{1961}]{Sandage:1961}
{Sandage} A.,  1961, {The Hubble Atlas of Galaxies}

\bibitem[\protect\citeauthoryear{{Sanders} \& {Binney}}{{Sanders} \&
  {Binney}}{2015}]{Sanders:2015}
{Sanders} J.~L.,  {Binney} J.,  2015, \mn@doi [\mnras] {10.1093/mnras/stv578},
  \href {http://adsabs.harvard.edu/abs/2015MNRAS.449.3479S} {449, 3479}

\bibitem[\protect\citeauthoryear{{Scannapieco}, {Gadotti}, {Jonsson}  \&
  {White}}{{Scannapieco} et~al.}{2010}]{Scannapieco:2010}
{Scannapieco} C.,  {Gadotti} D.~A.,  {Jonsson} P.,   {White} S.~D.~M.,  2010,
  \mn@doi [\mnras] {10.1111/j.1745-3933.2010.00900.x}, \href
  {http://adsabs.harvard.edu/abs/2010MNRAS.407L..41S} {407, L41}

\bibitem[\protect\citeauthoryear{{Scannapieco}, {White}, {Springel}  \&
  {Tissera}}{{Scannapieco} et~al.}{2011}]{Scannapieco:2011}
{Scannapieco} C.,  {White} S.~D.~M.,  {Springel} V.,   {Tissera} P.~B.,  2011,
  \mn@doi [\mnras] {10.1111/j.1365-2966.2011.19027.x}, \href
  {http://adsabs.harvard.edu/abs/2011MNRAS.417..154S} {417, 154}

\bibitem[\protect\citeauthoryear{{Schaye}, {Crain}, {Bower}  \& {et
  al.}}{{Schaye} et~al.}{2015}]{Schaye:2015}
{Schaye} J.,  {Crain} R.~A.,  {Bower} R.~G.,   {et al.} 2015, \mn@doi [\mnras]
  {10.1093/mnras/stu2058}, \href
  {http://adsabs.harvard.edu/abs/2015MNRAS.446..521S} {446, 521}

\bibitem[\protect\citeauthoryear{{Sch\"{o}lkopf}, {Smola}  \&
  {M\"{u}ller}}{{Sch\"{o}lkopf} et~al.}{1998}]{Scholkopf:1998}
{Sch\"{o}lkopf} B.,  {Smola} A.,   {M\"{u}ller} K.-R.,  1998, {Neural Comput.},
  10, 1299

\bibitem[\protect\citeauthoryear{{Searle} \& {Zinn}}{{Searle} \&
  {Zinn}}{1978}]{Searle:1978}
{Searle} L.,  {Zinn} R.,  1978, \mn@doi [\apj] {10.1086/156499}, \href
  {http://adsabs.harvard.edu/abs/1978ApJ...225..357S} {225, 357}

\bibitem[\protect\citeauthoryear{{Semelin} \& {Combes}}{{Semelin} \&
  {Combes}}{2002}]{Semelin:2002}
{Semelin} B.,  {Combes} F.,  2002, \mn@doi [\aap] {10.1051/0004-6361:20020336},
  \href {http://adsabs.harvard.edu/abs/2002A%26A...387...98S} {387, 98}

\bibitem[\protect\citeauthoryear{{Serna}, {Dom{\'{\i}}nguez-Tenreiro}  \&
  {S{\'a}iz}}{{Serna} et~al.}{2003}]{Serna:2003}
{Serna} A.,  {Dom{\'{\i}}nguez-Tenreiro} R.,   {S{\'a}iz} A.,  2003, \mn@doi
  [\apj] {10.1086/378629}, \href
  {http://adsabs.harvard.edu/abs/2003ApJ...597..878S} {597, 878}

\bibitem[\protect\citeauthoryear{{Shen}, {Wadsley}  \& {Stinson}}{{Shen}
  et~al.}{2010}]{Shen:2010}
{Shen} S.,  {Wadsley} J.,   {Stinson} G.,  2010, \mn@doi [\mnras]
  {10.1111/j.1365-2966.2010.17047.x}, \href
  {http://adsabs.harvard.edu/abs/2010MNRAS.407.1581S} {407, 1581}

\bibitem[\protect\citeauthoryear{{Sofue}, {Honma}  \& {Omodaka}}{{Sofue}
  et~al.}{2009}]{Sofue:2009}
{Sofue} Y.,  {Honma} M.,   {Omodaka} T.,  2009, \mn@doi [\pasj]
  {10.1093/pasj/61.2.227}, \href
  {http://adsabs.harvard.edu/abs/2009PASJ...61..227S} {61, 227}

\bibitem[\protect\citeauthoryear{{Staveley-Smith} \& {Davies}}{{Staveley-Smith}
  \& {Davies}}{1988}]{Staveley-Smith:1988}
{Staveley-Smith} L.,  {Davies} R.~D.,  1988, \mn@doi [\mnras]
  {10.1093/mnras/231.4.833}, \href
  {http://adsabs.harvard.edu/abs/1988MNRAS.231..833S} {231, 833}

\bibitem[\protect\citeauthoryear{{Stinson}, {Seth}, {Katz}, {Wadsley},
  {Governato}  \& {Quinn}}{{Stinson} et~al.}{2006}]{Stinson:2006}
{Stinson} G.,  {Seth} A.,  {Katz} N.,  {Wadsley} J.,  {Governato} F.,   {Quinn}
  T.,  2006, \mn@doi [\mnras] {10.1111/j.1365-2966.2006.11097.x}, \href
  {http://adsabs.harvard.edu/abs/2006MNRAS.373.1074S} {373, 1074}

\bibitem[\protect\citeauthoryear{{Stinson}, {Brook}, {Macci{\`o}}, {Wadsley},
  {Quinn}  \& {Couchman}}{{Stinson} et~al.}{2013}]{Stinson:2013a}
{Stinson} G.~S.,  {Brook} C.,  {Macci{\`o}} A.~V.,  {Wadsley} J.,  {Quinn}
  T.~R.,   {Couchman} H.~M.~P.,  2013, \mnras

\bibitem[\protect\citeauthoryear{{Thielemann}, {Nomoto}  \&
  {Yokoi}}{{Thielemann} et~al.}{1986}]{Thielemann:1986}
{Thielemann} F.-K.,  {Nomoto} K.,   {Yokoi} K.,  1986, \aap, \href
  {http://adsabs.harvard.edu/abs/1986A%26A...158...17T} {158, 17}

\bibitem[\protect\citeauthoryear{{Tully} \& {Fisher}}{{Tully} \&
  {Fisher}}{1977}]{Tully:1977}
{Tully} R.~B.,  {Fisher} J.~R.,  1977, \aap, \href
  {http://adsabs.harvard.edu/abs/1977A%26A....54..661T} {54, 661}

\bibitem[\protect\citeauthoryear{{Wadsley}, {Stadel}  \& {Quinn}}{{Wadsley}
  et~al.}{2004}]{Wadsley:2004}
{Wadsley} J.~W.,  {Stadel} J.,   {Quinn} T.,  2004, \mn@doi [\na]
  {10.1016/j.newast.2003.08.004}, \href
  {http://adsabs.harvard.edu/abs/2004NewA....9..137W} {9, 137}

\bibitem[\protect\citeauthoryear{{Wadsley}, {Veeravalli}  \&
  {Couchman}}{{Wadsley} et~al.}{2008}]{Wadsley:2008}
{Wadsley} J.~W.,  {Veeravalli} G.,   {Couchman} H.~M.~P.,  2008, \mn@doi
  [\mnras] {10.1111/j.1365-2966.2008.13260.x}, \href
  {http://adsabs.harvard.edu/abs/2008MNRAS.387..427W} {387, 427}

\bibitem[\protect\citeauthoryear{{Wadsley}, {Keller}  \& {Quinn}}{{Wadsley}
  et~al.}{2017}]{Wadsley:2017}
{Wadsley} J.~W.,  {Keller} B.~W.,   {Quinn} T.~R.,  2017, \mn@doi [\mnras]
  {10.1093/mnras/stx1643}, \href
  {http://adsabs.harvard.edu/abs/2017MNRAS.471.2357W} {471, 2357}

\bibitem[\protect\citeauthoryear{{Walt}, {Colbert}  \& {Varoquaux}}{{Walt}
  et~al.}{2011}]{Walt:2011}
{Walt} S. v.~d.,  {Colbert} S.~C.,   {Varoquaux} G.,  2011, Computing in
  Science and Engineering, 13, 22

\bibitem[\protect\citeauthoryear{{Wang}, {Dutton}, {Stinson}, {Macci{\`o}},
  {Penzo}, {Kang}, {Keller}  \& {Wadsley}}{{Wang} et~al.}{2015}]{Wang:2015}
{Wang} L.,  {Dutton} A.~A.,  {Stinson} G.~S.,  {Macci{\`o}} A.~V.,  {Penzo} C.,
   {Kang} X.,  {Keller} B.~W.,   {Wadsley} J.,  2015, \mn@doi [\mnras]
  {10.1093/mnras/stv1937}, \href
  {http://adsabs.harvard.edu/abs/2015MNRAS.454...83W} {454, 83}

\bibitem[\protect\citeauthoryear{{White}}{{White}}{1984}]{White:1984}
{White} S.~D.~M.,  1984, \mn@doi [\apj] {10.1086/162573}, \href
  {http://adsabs.harvard.edu/abs/1984ApJ...286...38W} {286, 38}

\bibitem[\protect\citeauthoryear{{Woosley} \& {Weaver}}{{Woosley} \&
  {Weaver}}{1995}]{Woosley:1995}
{Woosley} S.~E.,  {Weaver} T.~A.,  1995, \mn@doi [\apjs] {10.1086/192237},
  \href {http://adsabs.harvard.edu/abs/1995ApJS..101..181W} {101, 181}

\bibitem[\protect\citeauthoryear{{Yoachim} \& {Dalcanton}}{{Yoachim} \&
  {Dalcanton}}{2006}]{Yoachim:2006}
{Yoachim} P.,  {Dalcanton} J.~J.,  2006, \mn@doi [\aj] {10.1086/497970}, \href
  {http://adsabs.harvard.edu/abs/2006AJ....131..226Y} {131, 226}

\bibitem[\protect\citeauthoryear{{Yoachim} \& {Dalcanton}}{{Yoachim} \&
  {Dalcanton}}{2008}]{Yoachim:2008b}
{Yoachim} P.,  {Dalcanton} J.~J.,  2008, \mn@doi [\apj] {10.1086/589553}, \href
  {http://adsabs.harvard.edu/abs/2008ApJ...682.1004Y} {682, 1004}

\bibitem[\protect\citeauthoryear{{van Dokkum}, {Leja}, {Nelson}  \& {et
  al.}}{{van Dokkum} et~al.}{2013}]{vanDokkum:2013}
{van Dokkum} P.~G.,  {Leja} J.,  {Nelson} E.~J.,   {et al.} 2013, \mn@doi
  [\apjl] {10.1088/2041-8205/771/2/L35}, \href
  {http://adsabs.harvard.edu/abs/2013ApJ...771L..35V} {771, L35}

\bibitem[\protect\citeauthoryear{{van Dokkum}, {Abraham}, {Merritt}, {Zhang},
  {Geha}  \& {Conroy}}{{van Dokkum} et~al.}{2015}]{vanDokkum:2015}
{van Dokkum} P.~G.,  {Abraham} R.,  {Merritt} A.,  {Zhang} J.,  {Geha} M.,
  {Conroy} C.,  2015, \mn@doi [\apjl] {10.1088/2041-8205/798/2/L45}, \href
  {http://adsabs.harvard.edu/abs/2015ApJ...798L..45V} {798, L45}

\bibitem[\protect\citeauthoryear{{van der Wel}, {Bell}, {H{\"a}ussler}  \& {et.
  al}}{{van der Wel} et~al.}{2012}]{vanderWel:2012}
{van der Wel} A.,  {Bell} E.~F.,  {H{\"a}ussler} B.,   {et. al} 2012, \mn@doi
  [\apjs] {10.1088/0067-0049/203/2/24}, \href
  {http://esoads.eso.org/abs/2012ApJS..203...24V} {203, 24}

\makeatother
\end{thebibliography}

\end{document}